\documentclass[aps,prd,preprint,showkeys,showpacs,unsortedaddress,superscriptaddress]{revtex4-1}
\usepackage{amsmath, amsfonts, amssymb, mathrsfs,float}
\usepackage{graphicx}% Include figure files
\usepackage{bm,color}% bold math
\newcommand{\nn}{\nonumber}

\newcommand{\sss}{\scriptscriptstyle}
\allowdisplaybreaks

\begin{document}
%\preprint{APS/123-QED}

\title{Frequency shift of light in Kerr spacetime}
\author{Chunhua Jiang}%
\email{jiangchunhua@usc.edu.cn}
\affiliation{School of Mathematics and Physics, University of South China, Hengyang, 421001, China}
\affiliation{Hunan Key Laboratory of Mathematical Modeling and Scientific Computing, University of South China, Hengyang, 421001, China}
\author{Bo Yang}
\affiliation{School of Mathematics and Physics, University of South China, Hengyang, 421001, China}
\affiliation{Purple Mountain Observatory, Chinese Academy of Sciences, Nanjing 210023, China}
\author{Guansheng He}
\affiliation{School of Mathematics and Physics, University of South China, Hengyang, 421001, China}
\affiliation{Purple Mountain Observatory, Chinese Academy of Sciences, Nanjing 210023, China}

\author{Wenbin Lin}
\email{To whom all correspondence should be addressed (lwb@usc.edu.cn).}
\affiliation{School of Mathematics and Physics, University of South China, Hengyang, 421001, China}
\affiliation{Hunan Key Laboratory of Mathematical Modeling and Scientific Computing, University of South China, Hengyang, 421001, China}
%\affiliation{School of Physical Science and Technology, Southwest Jiaotong University, Chengdu, 610031, China}
\date{\today}

\begin{abstract}
The frequency shift of light in the gravitational field generated by a rotating body is investigated. We consider the scenario in which both the light source and the observer are in motion. The frequency shift is calculated up to the second-order post-Minkowskian approximation via two different methods and the same result is achieved. The higher-order effects of the gravitational source's rotation on the frequency shift is obtained. Especially, when both the light source and the observer are located in the asymptotically flat region, an elegant formula is obtained, which can be easily used in the astronomical observations to determine the rotating gravitational source's mass and angular momentum.
%The frequency shift of a light ray emitted from a moving light source and traveling along null geodesics towards a moving observer is investigated in the presence of a weak gravitational background generated by a rotating body.
%By examining two conventional approaches, we derived consistent analytical expressions up to second-and-half-order post-Newtonian accuracy for calculating the frequency shift of light. These expressions not only enhance the accuracy of frequency shift computations but also contribute to a better understanding of light's behavior in a weak gravitational background.
%With the advancement of space technology, it has become imperative to account for the relativistic effects arising from the motion of both the light source and the observer in high-precision space missions. These findings are of utmost significance for accurately predicting and correcting frequency shifts in observational data.
\end{abstract}

\maketitle

\section{introduction}
The theory of general relativity, formulated by Einstein, serves as the prevailing framework for comprehending gravitational interactions. Experimental observations consistently validate the predictions set forth by this theory.
One of the key predictions arising from general relativity is the phenomenon of gravitational redshift in light rays. This occurrence not only substantiates the credibility of the principle of equivalence, which forms the foundation of general relativity, but also serves as substantial evidence supporting the theory's veracity.

The first reported confirmation of gravitational redshift originated from the measurement of the apparent radial velocity of Sirius B, where the observed motion deviated from the expected motion within its 50-year orbit\cite{Adams1925}.
Pound-Rebka laboratory experiment\cite{Pound-Rebka1959} of 1960s was the first to experimentally verify the gravitational redshift in nuclear resonance, i.e., the M\"{o}ssbauer resonant effect, which measured the tiny frequency shift of photons on the surface of the Earth which fall down a tower.
A fundamental examination of gravitational redshift, known as the Gravity Probe A, was conducted by Vessot et al.\cite{Vessot1979,Vessot1980,Vessot1989}. This experiment involved measuring the frequency shift of a hydrogen maser on a spacecraft compared to a similar maser on Earth.
Currently, atom interferometer tests\cite{Muller2010,Poli2011} provide gravitational redshift measurements with a precision that is approximately 10,000 times greater than traditional clock-based tests\cite{Vessot1979,Vessot1980,Vessot1989}.

The frequency shift of a light ray is a complex phenomenon resulting from a combination of both kinematical and gravitational effects. The kinematical contribution is known as the Doppler effect, while the gravitational contribution arises from the influence of gravity. In the realm of special relativity, the Doppler effect between the emission and reception of light from moving sources is well understood.
In 1966, the GEOS-1 satellite was utilized to observe the relativistic Doppler effects, providing empirical evidence for these kinematical contributions\cite{Jenkins1969}. However, the gravitational frequency shift, in addition to the kinematical factor, is governed by the curvature of spacetime, as described by the principles of general relativity.
Understanding the interplay between these two effects is crucial for accurately interpreting and predicting the observed frequency shifts of light rays in various gravitational and kinematical scenarios.
In the Schwarzschild spacetime, the gravitational frequency shift of a photon emitted at a radial position $r_1$ and received at a different radial position $r_2$ is described by the following equation\cite{Ruffini1994}:
\begin{equation}
  \frac{\nu_1}{\nu_2}=\frac{\sqrt{1-2GM/r_2}}{\sqrt{1-2GM/r_1}}~,
\end{equation}
where $\nu_1$ and $\nu_2$ are the measured frequencies at the photon's emission and reception.
Krisher\cite{Krisher1993} has derived the gravitational frequency shift effect to order $c^{-4}$ ($c$ represents the velocity of light) within the framework of the parametrized post-Newtonian (PN) formalism for analyzing metric theories of gravity.
Kopeikin and Sch\"{a}fer\cite{Kopeikin1999} had derived the exact analytical expression for the gravitational shift of electromagnetic frequency caused by the gravitational fields of arbitrary moving bodies.
Dubey and Sen\cite{Sen2015a,Sen2015b} have derived the gravitational frequency shift expressions for a rotating body and a charged rotating body using the Kerr geometry and Kerr-Newman geometry, respectively.
Deng\cite{Deng2016} has obtained an expression for gravitational frequency shift within the framework of a second parametrized PN formalism in the solar system barycentric reference system.
Qin and Shao\cite{qin2017} have deduced the general post-Minkowskian (PM) expansion of the phase function from the eikonal function and determined the frequency shift up to the order of $c^{-3}$.
Recently, Kuntz and Leyde\cite{Kuntz2023} showed that the presence of the redshift/Doppler effect at 1PN order could be visible in the waveform of a binary situated close to a supermassive black hole.
In this paper, we discuss the frequency shift of a light ray emitted from a moving source and received by a moving observer, considering the weak gravitational background generated by a rotating compact body, up to the order of 2PM.

The structure of this paper is organized as follows. In Section \ref{sec:2nd}, we present the 2PM solutions for the trajectory and velocity of light traveling in the gravitational field of a Kerr black hole. Section \ref{sec:3rd} is devoted to derive the formulas for the frequency shift of a light ray emitted from an orbiting source and received by a moving observer, up to 2PM accuracy. These formulas are obtained by utilizing the relationship between the frequency of the light ray emitted from an orbiting source and the frequency of the light received by a moving observer. In Section \ref{sec:frequencyEnergy}, we present the same formulas, but the frequency shift of the light ray is expressed in terms of the relationship between the energy of the light ray emitted from an orbiting source and the energy of the light received by a moving observer. Some special scenarios are considered in Section \ref{sec:somecases}. Finally, summary is given in Section \ref{sec:summary}.
We use the geometrized units (the gravitational constant $G$ and the light speed $c$ in vacuum are set as $1$). The metric signature is given as ($-+++$). Greek indices run from 0--3 and Latin indices from 1--3. The Einstein convention on repeat indices is used here for the expressions like $a^i b^i$ and $A^\mu B_\mu$.

\section{The 2PM solution to the light propagation in Kerr spacetime\label{sec:2nd}}

The Kerr field is an exact solution of the Einstein equations that describes the exterior gravitational field of a constantly rotating source with mass $m$ and angular momentum per unit mass $a$. In the harmonic coordinates, the metric describing the Kerr field, in 2PM approximations, can be written as \cite{JiangLin2014,LinJiang2014}
\begin{eqnarray}\label{eq:metric-2PM}
g_{00} &=& -1 + \frac{2m}{r} - \frac{2m^2}{r^2}~, \nn\\
g_{0i} &=& \frac{2\epsilon_{ijk} x^j J^k}{r^3}~,\\
g_{ij} &=& \bigg(1 + \frac{2m}{r} + \frac{m^2}{r^2}\bigg)\delta_{ij} + \frac{m^2}{r^4}x^i x^j~, \nn
\end{eqnarray}
where $r\equiv |\bm{x}|$ with $\bm{x}\equiv(x, y, z)$ denoting the position vector of the field point. $\bm{J} = am\bm{e}_{z}$, with $\bm{e}_{z}$ being the unit vector of $z$-axis, represents the angular momentum of the gravitational source. The mass and angular momentum of the rotating source are restricted by the relation $m\geq a$ to ensure there is no naked singularity. $\epsilon_{ijk}$ is Levi-Civita symbol.

The motion of photon in the gravitational field can be described by the following equations~\cite{Weinberg1972,Will1981,KopeikinEfroimskyKaplan2012}
\begin{eqnarray}
  &&g_{00} + 2g_{0i}\frac{dx^i}{dt} + g_{ij}\frac{dx^i}{dt}\frac{dx^j}{dt} = 0~, \label{eq:Nullcondition}\\
  &&\frac{d^2x^i}{dt^2} + \Gamma^i_{00} +2\Gamma^i_{0j}\frac{dx^j}{dt} +\Gamma^i_{jk}\frac{dx^j}{dt}\frac{dx^k}{dt} - \left(\Gamma^0_{00}+ 2\Gamma^0_{0j}\frac{dx^j}{dt} +\Gamma^0_{jk}\frac{dx^j}{dt}\frac{dx^k}{dt}\right)\frac{dx^i}{dt}=0~,\label{eq:geodesicEq}
\end{eqnarray}
where $\Gamma_{\alpha\beta}^\mu$ is Christoffel symbol with
\begin{equation*}
  \Gamma_{\alpha\beta}^\mu = \frac{1}{2}g^{\rho\mu}\left(\frac{\partial g_{\rho\beta}}{\partial x^\alpha}+\frac{\partial g_{\rho\alpha}}{\partial x^\beta} - \frac{\partial g_{\alpha\beta}}{\partial x^\rho}\right)~.
\end{equation*}

Assuming a photon is emitted at the coordinate time $t_{\rm s}$ from a point $\bm{x}_{\rm s}$ with an initial direction unit vector $\bm{n}_{\rm s}$, we have previously derived the 2PM solution for light propagation in the Kerr-Newman spacetime in our paper~\cite{JiangLin2018}. Following a similar approach, we can solve Eqs. \eqref{eq:Nullcondition} and \eqref{eq:geodesicEq} iteratively and obtain the 2PM trajectory in the gravitational field of a Kerr black hole as follows:
\begin{equation}\label{eq:trajectory-2PM-general}
  \bm{x} = \bm{x}_{\sss\rm 0PM} + \bm{x}_{\sss\rm 1PM} + \bm{x}_{\sss\rm 2PM}~,
\end{equation}
where
\begin{equation}\label{eq:trajectory-0PM}
  \bm{x}_{\sss\rm 0PM} = \bm{x}_{\rm s} + \bm{n}_{\rm s}(t -t_{\rm s})~,
\end{equation}
\begin{equation}\label{eq:trajectory-1PM}
  \bm{x}_{\sss\rm 1PM} = -2m\bm{n}_{\rm s}\ln{\frac{r+\bm{n}_{\rm s}\cdot\bm{x}}{r_{\rm s}+\bm{n}_{\rm s}\cdot\bm{x}_{\rm s}}} - \frac{2m\bm{b}}{b^2}\bigg(r - \frac{\bm{x}_{\rm s}\cdot\bm{x}}{r_{\rm s}}\bigg)~,
\end{equation}
\begin{eqnarray}
\bm{x}_{\sss\rm 2PM} \!&=& \!\bm{n}_{\rm s}\bigg\{\!\frac{15m^2}{4b}\!\bigg(\!\!\arccos\!{\frac{\bm{n}_{\rm s}\!\cdot\!\bm{x}\!}{r}}\!-\!\arccos\!{\frac{\bm{n}_{\rm s}\!\cdot\!\bm{x}_{\rm s}\!}{r_{\rm s}}}\!\bigg) \!+\!\frac{m^2}{4}\!\bigg(\!\frac{\bm{n}_{\rm s}\!\cdot\!\bm{x}}{r^2} \!-\!\frac{\bm{n}_{\rm s}\!\cdot\!\bm{x}_{\rm s}}{r_{\rm s}^2}\!\bigg) \!+\! \frac{2m^2(t\!-\!t_{\rm s})}{r_{\rm s}^2} \nn\\
&&\hskip 0.5cm\!-\frac{4m^2 r}{b^2}\!\bigg(\!1\!-\!\frac{\bm{x}\!\cdot\!\bm{x}_{\rm s}}{rr_{\rm s}}\!\bigg)\!-\!\frac{2(\bm{n}_{\rm s}\!\times\!\bm{b})\!\cdot\!\bm{J}}{b^2}\!\bigg(\!\frac{\bm{n}_{\rm s}\!\cdot\!\bm{x}}{r} \!-\! \frac{\bm{n}_{\rm s}\!\cdot\!\bm{x}_{\rm s}}{r_{\rm s}}\!\bigg)\!\bigg\} \nn\\
&&\hskip -0.25cm +\bm{b}\bigg\{\!\frac{15m^2(\bm{n}_{\rm s}\!\cdot\!\bm{x})}{4b^3}\!\bigg(\!\!\arccos\!{\frac{\bm{n}_{\rm s}\!\cdot\!\bm{x}}{r}}\!-\!\arccos\!{\frac{\bm{n}_{\rm s}\!\cdot\!\bm{x}_{\rm s}}{r_{\rm s}}}\!\!\bigg) \!-\!\frac{m^2}{b^2}\!\bigg(\!\frac{15}{4}\!-\!\frac{8r}{r_{\rm s}}\!+\!\frac{17\bm{x}_{\rm s}\!\cdot\!\bm{x}}{4r_{\rm s}^2}\!\bigg) \nn\\
&&\hskip .5cm+\frac{m^2\bm{x}_{\rm s}\cdot\bm{x}}{2r_{\rm s}^4} \!-\!\frac{3m^2}{4r_{\rm s}^2}\!+\!\frac{m^2}{4r^2} \!-\! \frac{2r(\bm{n}_{\rm s}\!\times\!\bm{b})\!\cdot\!\bm{J}}{b^4}\!\bigg(\!1\!-\!\frac{\bm{x}_{\rm s}\!\cdot\!\bm{x}}{rr_{\rm s}}\!\bigg)\!\bigg\} \nn\\
&&\hskip -0.25cm +\bm{n}_{\rm s}\!\times\!\bm{b}\bigg\{\!\frac{2\bm{n}_{\rm s}\!\cdot\!\bm{J}}{b^2}\!\bigg[\!\frac{\bm{n}_{\rm s}\!\cdot\!\bm{x}}{r}\!-\!\frac{\!\bm{n}_{\rm s}\!\cdot\!\bm{x}_{\rm s}}{r_{\rm s}}\!-\!\frac{b^2(t\!-\!t_{\rm s})}{r_{\rm s}^3}\!\bigg]\!+\!\frac{2\bm{b}\!\cdot\!\bm{J}}{b^2}\!\bigg[\!\frac{1}{r}\!-\!\frac{2}{\!r_{\rm s}\!}\!-\!\frac{\!r\!}{b^2}\!+\!\frac{\bm{x}_{\rm s}\!\cdot\!\bm{x}\!}{r_{\rm s}}\!\bigg(\!\frac{1}{b^2}\!+\!\frac{1}{r_{\rm s}^2}\!\bigg)\!\bigg]\!\bigg\}~,  \label{eq:trajectory-2PM}
\end{eqnarray}
and $r_{\rm s} = |\bm{x}_{\rm s}|$, $\bm{x}_{\sss\rm 0PM}$ denotes the solution of light trajectory when the gravitational field absent, $\bm{x}_{\sss\rm 1PM}$ and $\bm{x}_{\sss\rm 2PM}$ represent the 1PM and 2PM corrections to the solution in the Minkowskian geometry, respectively. $\bm{b} \equiv  \bm{n}_{\rm s}\times(\bm{x}_{\sss\rm e}\times\bm{n}_{\rm s})$, and $b \equiv |\bm{b}|$ is well-known as the impact parameter \cite{Weinberg1972}.

\section{frequency shift in terms of frequency}\label{sec:3rd}
\begin{figure}[b]
  \centering
  \includegraphics[width=17cm]{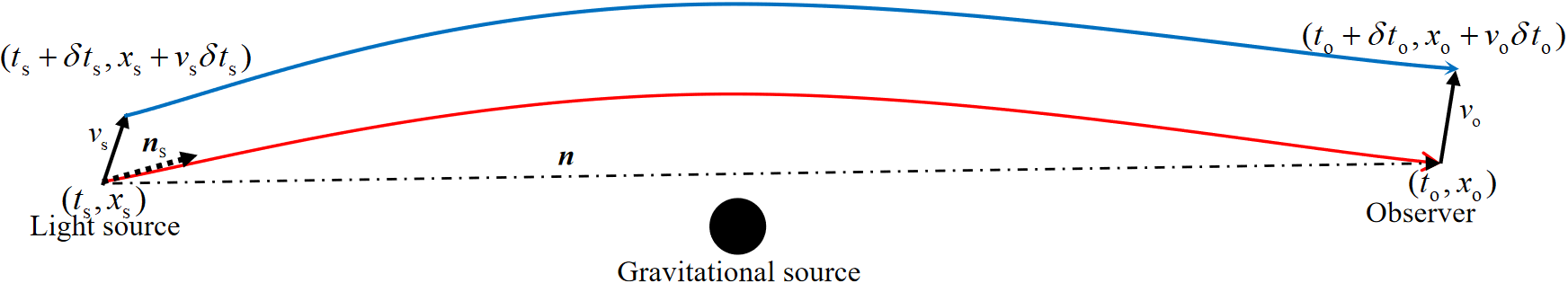}\\
  \caption{Spacetime sketch of the configuration. The gravitational source is stationary, the light source and the observer move with velocity $\bm{v}_{\sss\rm s}$ and $\bm{v}_{\sss\rm o}$, respectively. The red solid line represents the geodesic of the light-ray emitted from the light source at the coordinate time $t_{\rm s}$ from the point $\bm{x}_{\rm s}$ and received the observer at the coordinate time $t_{\rm o}$ at the point $\bm{x}_{\rm o}$. The blue solid line denotes the light geodesics emitted at the coordinate time $t_{\rm s} + \delta t_{\rm s}$ and received at coordinate time $t_{\rm o}+\delta t_{\rm o}$ from the point $\bm{x}_{\rm o} + \bm{v}_{\rm o}\delta t_{\rm o}$ when the light source moves to the position $\bm{x}_{\rm s} + \bm{v}_{\rm s}\delta t_{\rm s}$. Here, $\bm{n}_{\rm s}$ is a direction unit vector of light propagation at the emission, and $\bm{n}=\frac{\bm{x}_{\rm o}-\bm{x}_{\rm s}}{|\bm{x}_{\rm o}-\bm{x}_{\rm s}|}$.}\label{fig1}
\end{figure}
Here we consider a thought experiment (see FIG.\ref{fig1}), photons travel along null geodesics from the emitter (geodesic particle), skim over the gravitational source (Kerr black hole), and is received by the observer(a far away located observer, the Earth for practical purpose). The light source moves with respect to the harmonic coordinate system of the gravitational source with 3-velocity $\bm{v}_{\sss\rm s}$, and emit electromagnetic radiation with frequency $\nu_{\rm s} = 1/\delta\tau_{\rm s}$, where $\delta\tau_{\rm s}$ is proper period of the electromagnetic radiation at the point of emission. The observer, moving with 3-velocity $\bm{v}_{\sss\rm o}$ with respect to the coordinate system, measures the frequency of the received electromagnetic radiation, $\nu_{\rm o} = 1/\delta\tau_{\rm o}$, where $\delta\tau_{\sss\rm o}$ is proper period of the electromagnetic wave at the point of reception. $\delta\tau_{\rm s}$ and $\delta\tau_{\rm o}$ are can be considered as infinitesimally small, in the paper, we just keep the linear terms about $\delta$, then one can consider that the emitter and the detector move in uniform motion. Furthermore, we may neglect the gravitational fields of the emitter and the detector. Therefore, the observed gravitational frequency shift \cite{Synge1971,Brumberg1972} is
\begin{equation}\label{eq:FrequecyShift}
  \frac{\nu_{\rm o}}{\nu_{\rm s}}-1=\frac{\delta\tau_{\rm s}}{\delta\tau_{\rm o}}-1 = \frac{\delta\tau_{\rm s}}{\delta t_{\rm s}}\frac{\delta t_{\rm s}}{\delta t_{\rm o}}\frac{\delta t_{\rm o}}{\delta\tau_{\rm o}}-1 = \frac{u^0(t_{\rm o}, \bm{x}_{\rm o})}{u^0(t_{\rm s}, \bm{x}_{\rm s})}\frac{\delta t_{\rm s}}{\delta t_{\rm o}}-1~,
\end{equation}
where $\delta t_{\rm s}$ and $\delta t_{\rm o}$ are coordinate periods of the electromagnetic radiation at the points of emission and reception, respectively. $\delta t_{\rm s}/\delta t_{\rm o}$ denotes the coordinate period ratio of the electromagnetic radiation at the points of emission and reception. Using proper time defined by $d\tau^2 = -g_{\mu\nu}dx^\mu dx^\nu$ and metric components \eqref{eq:metric-2PM}, we obtain, to 2PM accuracy,
\begin{equation}\label{eq:u^0}
u^0 =\gamma\Bigg[1+(2\gamma^2-1)\frac{m}{r}+\Big(1-\frac{13}{2}\gamma^2+6\gamma^4\Big)\frac{m^2}{r^2}+\frac{\gamma^2}{2}\Big(\bm{x}\cdot\frac{d\bm{x}}{dt}\Big)^2\frac{m^2}{r^4}+2\gamma^2\frac{(\bm{x}\times\bm{J})\cdot\frac{d\bm{x}}{dt}}{r^3}\Bigg]~,
\end{equation}
with $\gamma=(1\!-\!\frac{d\bm{x}}{dt}\!\cdot\!\frac{d\bm{x}}{dt})^{-1/2}$ being Lorentz factor. This expression can be used to calculate the ratio $u^0(t_{\rm o}, \bm{x}_{\rm o})/u^0(t_{\rm s}, \bm{x}_{\rm s})$ in \eqref{eq:FrequecyShift}. The result is
\begin{eqnarray}
\frac{u^0(t_{\rm o}, \bm{x}_{\rm o})}{u^0(t_{\rm s}, \bm{x}_{\rm s})} &=& \frac{\gamma_{\rm o}}{\gamma_{\rm s}} \Bigg[1+(1-2\gamma^2_{\rm s})\frac{m}{r_{\rm s}} - (1-2\gamma_{\rm o}^2)\frac{m}{r_{\rm o}} + \frac{m^2}{r_{\rm o}^2} - \frac{m^2}{r_{\rm o}r_{\rm s}} - \frac{13\gamma_{\rm o}^2}{2}\frac{m^2}{r_{\rm o}^2} + 2\gamma_{\rm o}^2\frac{m^2}{r_{\rm o}r_{\rm s}} \nn\\
&&+ 6\gamma_{\rm o}^4\frac{m^2}{r_{\rm o}^2} + \frac{5\gamma_{\rm s}^2}{2}\frac{m^2}{r_{\rm s}^2} + 2\gamma_{\rm s}^2\frac{m^2}{r_{\rm o}r_{\rm s}} - 4\gamma_{\rm o}^2\gamma_{\rm s}^2\frac{m^2}{r_{\rm o}r_{\rm s}} - 2\gamma_{\rm s}^4\frac{m^2}{r_{\rm s}^2} + \frac{\gamma_{\rm o}^2}{2}\frac{m^2(\bm{x}_{\rm o}\cdot\bm{v}_{\rm o})^2}{r_{\rm o}^4} \nn\\
&&-\frac{\gamma_{\rm s}^2}{2}\frac{m^2(\bm{x}_{\rm s}\cdot\bm{v}_{\rm s})^2}{r_{\rm s}^4} + 2\gamma_{\rm o}^2\frac{(\bm{x}_{\rm o}\times\bm{J})\cdot\bm{v}_{\rm o}}{r_{\rm o}^3} -2\gamma_{\rm s}^2\frac{(\bm{x}_{\rm s}\times\bm{J})\cdot\bm{v}_{\rm s}}{r_{\rm s}^3}\Bigg]~, \label{eq:coordinate_clock_ratio}
\end{eqnarray}
where $\gamma_{\rm o}=(1\!-\!\frac{d\bm{x}_{\rm o}}{dt_{\rm o}}\!\cdot\!\frac{d\bm{x}_{\rm o}}{dt_{\rm o}})^{-1/2}$ and $\gamma_{\rm s}=(1\!-\!\frac{d\bm{x}_{\rm s}}{dt_{\rm s}}\!\cdot\!\frac{d\bm{x}_{\rm s}}{dt_{\rm s}})^{-1/2}$.

In order to calculate $\delta t_{\rm s}/\delta t_{\rm o}$, we have to take into account the time delay for the points of emission and reception of electromagnetic radiation,
\begin{equation}\label{eq:TimeDelay_1}
  t_{\rm o} - t_{\rm s} = R - \bm{n}\cdot\bm{x}_{\sss\rm 1PM} - \bm{n}\cdot\bm{x}_{\sss\rm 2PM} + \frac{(\bm{n}\times\bm{x}_{\sss\rm 1PM})^2}{2R}~,
\end{equation}
where $R = |\bm{x}_{\rm o} - \bm{x}_{\rm s}|$, $\bm{n} = \frac{\bm{x}_{\rm o}-\bm{x}_{\rm s}}{R}$, the relationship of $\bm{n}_{\rm s}$ and $\bm{n}$ is
\begin{equation}\label{eq:n&ne}
  \bm{n}_{\rm s} = \bm{n} \!+\! \frac{\bm{n}\!\times\!(\!\bm{n}\!\times\!\bm{x}_{\sss\rm 1PM}\!)}{R} \!+\! \frac{\bm{n}\!\times\!(\!\bm{n}\!\times\!\bm{x}_{\sss\rm 2PM}\!)}{R} \!-\! \frac{3(\!\bm{n}\!\times\!\bm{x}_{\sss\rm 1PM}\!)^2}{2R^2}\bm{n} \!+\! \frac{\bm{x}_{\sss\rm 1PM}\!\times\!(\!\bm{n}\!\times\!\bm{x}_{\sss\rm 1PM}\!)}{R^2}~.
\end{equation}

Similarly, the time delay for two events $(t_{\rm s}+\delta t_{\rm s}, \bm{x}_{\rm s} + \bm{v}_{\rm s}\delta t_{\rm s})$ and $(t_{\rm o}+\delta t_{\rm o}, \bm{x}_{\rm o} + \bm{v}_{\rm o}\delta t_{\rm o})$ is
\begin{equation}\label{eq:TimeDelay_2}
  \tilde{t}_{\rm o}-\tilde{t}_{\rm s} = \tilde{R} - \tilde{\bm{n}}\cdot\tilde{\bm{x}}_{\sss\rm 1PM} - \tilde{\bm{n}}\cdot\tilde{\bm{x}}_{\sss\rm 2PM} + \frac{(\tilde{\bm{n}}\times\tilde{\bm{x}}_{\sss\rm 1PM})^2}{2\tilde{R}}~,
\end{equation}
where
\begin{equation}\label{eq:trajectory-1PM-tilde}
  \tilde{\bm{x}}_{\sss\rm 1PM} = -2m\tilde{\bm{n}}_{\rm s}\ln{\frac{\tilde{r}_{\rm o}+\tilde{\bm{n}}_{\rm s}\cdot\tilde{\bm{x}}_{\rm o}}{\tilde{r}_{\rm s}+\tilde{\bm{n}}_{\rm s}\cdot\tilde{\bm{x}}_{\rm s}}} - \frac{2m\tilde{\bm{b}}}{\tilde{b}^2}\bigg(\tilde{r}_{\rm o} - \frac{\tilde{\bm{x}}_{\rm s}\cdot\tilde{\bm{x}}_{\rm o}}{\tilde{r}_{\rm s}}\bigg)~,
\end{equation}
\begin{eqnarray}
\tilde{\bm{x}}_{\sss\rm 2PM} \!&=& \!\tilde{\bm{n}}_{\rm s}\bigg\{\!\frac{15m^2}{4\tilde{b}}\!\bigg(\!\!\arccos\!{\frac{\tilde{\bm{n}}_{\rm s}\!\cdot\!\tilde{\bm{x}}\!}{\tilde{r}_{\rm o}}}\!-\!\arccos\!{\frac{\tilde{\bm{n}}_{\rm s}\!\cdot\!\tilde{\bm{x}}_{\rm s}\!}{\tilde{r}_{\rm s}}}\!\bigg) \!+\!\frac{m^2}{4}\!\bigg(\!\frac{\tilde{\bm{n}}_{\rm s}\!\cdot\!\tilde{\bm{x}}_{\rm o}}{\tilde{r}^2_{\rm o}} \!-\!\frac{\tilde{\bm{n}}_{\rm s}\!\cdot\!\tilde{\bm{x}}_{\rm s}}{\tilde{r}_{\rm s}^2}\!\bigg) \!+\! \frac{2m^2(\tilde{t}_{\rm o}\!-\!\tilde{t}_{\rm s})}{\tilde{r}_{\rm s}^2} \nn\\
&&\hskip0.5cm\!-\frac{4m^2 \tilde{r}_{\rm o}}{\tilde{b}^2}\!\bigg(\!1\!-\!\frac{\tilde{\bm{x}}_{\rm o}\!\cdot\!\tilde{\bm{x}}_{\rm s}}{\tilde{r}_{\rm o}\tilde{r}_{\rm s}}\!\bigg) \!-\!\frac{2(\tilde{\bm{n}}_{\rm s}\!\times\!\tilde{\bm{b}})\!\cdot\!\bm{J}}{\tilde{b}^2}\!\bigg(\!\frac{\tilde{\bm{n}}_{\rm s}\!\cdot\!\tilde{\bm{x}}_{\rm o}}{\tilde{r}_{\rm o}} \!-\! \frac{\tilde{\bm{n}}_{\rm s}\!\cdot\!\tilde{\bm{x}}_{\rm s}}{\tilde{r}_{\rm s}}\!\bigg)\!\bigg\} \nn\\
&&\hskip -1cm  +\tilde{\bm{b}}\bigg\{\!\frac{15m^2(\tilde{\bm{n}}_{\rm s}\!\cdot\!\tilde{\bm{x}}_{\rm o})}{4\tilde{b}^3}\!\bigg(\!\!\arccos\!{\frac{\tilde{\bm{n}}_{\rm s}\!\cdot\!\tilde{\bm{x}}_{\rm o}}{\tilde{r}_{\rm o}}}\!-\!\arccos\!{\frac{\tilde{\bm{n}}_{\rm s}\!\cdot\!\tilde{\bm{x}}_{\rm s}}{\tilde{r}_{\rm s}}}\!\!\bigg) \!-\!\frac{m^2}{\tilde{b}^2}\!\bigg(\!\frac{15}{4}\!-\!\frac{8\tilde{r}_{\rm o}}{\tilde{r}_{\rm s}}\!+\!\frac{17\tilde{\bm{x}}_{\rm s}\!\cdot\!\tilde{\bm{x}}_{\rm o}}{4\tilde{r}_{\rm s}^2}\!\bigg) \nn\\
&&\hskip-0.5cm +\frac{m^2\tilde{\bm{x}}_{\rm s}\!\cdot\!\tilde{\bm{x}}_{\rm o}}{2\tilde{r}_{\rm s}^4} \!-\!\frac{3m^2}{4\tilde{r}_{\rm s}^2}\!+\!\frac{m^2}{4\tilde{r}^2_{\rm o}} \!-\! \frac{2\tilde{r}(\tilde{\bm{n}}_{\rm s}\!\times\!\tilde{\bm{b}})\!\cdot\!\bm{J}}{\tilde{b}^4}\!\bigg(\!1\!-\!\frac{\tilde{\bm{x}}_{\rm s}\!\cdot\!\tilde{\bm{x}}_{\rm o}}{\tilde{r}_{\rm o}\tilde{r}_{\rm s}}\!\bigg)\!\bigg\} \nn\\
&&\hskip -1cm +\tilde{\bm{n}}_{\rm s}\!\times\!\tilde{\bm{b}}\bigg\{\!\frac{2\tilde{\bm{n}}_{\rm s}\!\cdot\!\bm{J}}{\tilde{b}^2}\!\bigg[\!\frac{\tilde{\bm{n}}_{\rm s}\!\cdot\!\tilde{\bm{x}}_{\rm o}}{\tilde{r}_{\rm o}}\!-\!\frac{\!\tilde{\bm{n}}_{\rm s}\!\cdot\!\tilde{\bm{x}}_{\rm s}}{\tilde{r}_{\rm s}}\!-\!\frac{\tilde{b}^2(\tilde{t}_{\rm o}\!-\!\tilde{t}_{\rm s})}{\tilde{r}_{\rm s}^3}\!\bigg]\!+\!\frac{2\tilde{\bm{b}}\!\cdot\!\bm{J}}{\tilde{b}^2}\!\bigg[\!\frac{1}{\tilde{r}_{\rm o}}\!-\!\frac{2}{\!\tilde{r}_{\rm s}\!}\!-\!\frac{\!\tilde{r}_{\rm o}\!}{\tilde{b}^2}\!+\!\frac{\tilde{\bm{x}}_{\rm s}\!\cdot\!\tilde{\bm{x}}_{\rm o}}{\tilde{r}_{\rm s}}\!\bigg(\!\frac{1}{\tilde{b}^2}\!+\!\frac{1}{\tilde{r}_{\rm s}^2}\!\bigg)\!\!\bigg]\!\!\bigg\},  \label{eq:trajectory-2PM-tilde}
\end{eqnarray}
\begin{equation}
  \tilde{\bm{n}}_{\rm s} = \tilde{\bm{n}}\!+\!\frac{\tilde{\bm{n}}\!\times\!(\!\tilde{\bm{n}}\!\times\!\tilde{\bm{x}}_{\sss\rm 1PM}\!)}{\tilde{R}} \!+\! \frac{\tilde{\bm{n}}\!\times\!(\!\tilde{\bm{n}}\!\times\!\tilde{\bm{x}}_{\sss\rm 2PM}\!)}{\tilde{R}} \!-\! \frac{3(\!\tilde{\bm{n}}\!\times\!\tilde{\bm{x}}_{\sss\rm 1PM}\!)^2}{2\tilde{R}^2}\tilde{\bm{n}} \!+\! \frac{\tilde{\bm{x}}_{\sss\rm 1PM}\!\times\!(\!\tilde{\bm{n}}\!\times\!\tilde{\bm{x}}_{\sss\rm 1PM}\!)}{\tilde{R}^2}~,
\end{equation}
and $\tilde{t}_{\rm o} = t_{\rm o} + \delta t_{\rm o}$, $\tilde{t}_{\rm s} = t_{\rm s} + \delta t_{\rm s}$, $\tilde{\bm{x}}_{\rm o} = \bm{x}_{\rm o} + \bm{v}_{\rm o}\delta t_{\rm o}$, $\tilde{r}_{\rm o} = |\tilde{\bm{x}}_{\rm o}|$, $\tilde{\bm{x}}_{\rm s} = \bm{x}_{\rm s} + \bm{v}_{\rm s}\delta t_{\rm s}$, $\tilde{r}_{\rm s} = |\tilde{\bm{x}}_{\rm s}|$, $\tilde{R} = |\tilde{\bm{x}}_{\rm o} - \tilde{\bm{x}}_{\rm s}|$, $\tilde{\bm{n}} = \frac{\tilde{\bm{x}}_{\rm o} - \tilde{\bm{x}}_{\rm s}}{\tilde{R}}$, $\tilde{\bm{b}} = \tilde{\bm{n}}_{\rm s}\times(\tilde{\bm{x}}_{\rm s}\times\tilde{\bm{n}}_{\rm s})$, $\tilde{b} = |\tilde{\bm{b}}|$.
Keeping the 2PM terms and the linear terms of $\delta$, Eq. \eqref{eq:TimeDelay_2} takes away Eq. \eqref{eq:TimeDelay_1} to get
\begin{eqnarray}
  \delta t_{\rm o} \!-\! \delta t_{\rm s} &=& \bigg[1 - \frac{(\bm{n}\times\bm{x}_{\sss\rm 1PM})^2}{2R^2}\bigg]\delta R - (\bm{x}_{\sss\rm1PM}+\bm{x}_{\sss\rm2PM})\cdot\delta\bm{n} - \bm{n}\cdot(\delta\bm{x}_{\sss\rm1PM}+\delta\bm{x}_{\sss\rm2PM}) \nn\\
  &&+ \frac{\bm{n}\times\bm{x}_{\sss\rm1PM}}{R}\cdot(\bm{n}\times\delta\bm{x}_{\sss\rm1PM}+\delta\bm{n}\times\bm{x}_{\sss\rm1PM})~,\label{eq:radio-time}
\end{eqnarray}
where $\delta R=\tilde{R}-R$, $\delta n=\tilde{\bm{n}}-\bm{n}$, $\delta\bm{x}_{\sss\rm1PM}=\tilde{\bm{x}}_{\sss\rm1PM}-\bm{x}_{\sss\rm1PM}$ and $\delta\bm{x}_{\sss\rm2PM}=\tilde{\bm{x}}_{\sss\rm2PM}-\bm{x}_{\sss\rm2PM}$.

After straightforward and tedious calculation, we simplify Eq. \eqref{eq:radio-time} in the 2PM approximation, is
\begin{equation}\label{eq:derivative-time}
\frac{\delta t_{\rm s}}{\delta t_{\rm o}}=\frac{A}{B}~,
\end{equation}
where
\begin{eqnarray*}
A &=& 1\!-\!\!\bigg[1\!+\!\frac{2m}{r_{\rm o}}\!+\!\frac{4m^2}{r^2_{\rm o}}\!-\!\frac{4m^2}{b^2}\!+\!\frac{2m^2}{r_{\rm s}^2}\!-\!\frac{m^2b^2}{2r^4_{\rm o}}\!+\!\frac{4m^2}{b^2}\frac{\bm{n}_{\rm s}\!\cdot\!\bm{x}_{\rm s}}{r_{\rm s}}\frac{\bm{n}_{\rm s}\!\cdot\!\bm{x}_{\rm o}}{r_{\rm o}}\!+\!\frac{2(\bm{n}_{\rm s}\!\times\!\bm{b})\cdot\bm{J}}{r^3_{\rm o}}\bigg]\!\bm{n}_{\rm s}\!\cdot\bm{v}_{\rm o} \\
&&\hskip-.25cm+\bigg\{\!\frac{2m}{b^2}\!\Big(\!\frac{\bm{n}_{\rm s}\!\cdot\!\bm{x}_{\rm o}}{r_{\rm o}}\!-\!\frac{\bm{n}_{\rm s}\!\cdot\!\bm{x}_{\rm s}}{r_{\rm s}}\!\Big)\!-\!\frac{15m^2}{4b^3}\!\Big(\!\!\arccos{\frac{\bm{n}_{\rm s}\!\cdot\!\bm{x}_{\rm o}}{r_{\rm o}}}\!-\!\arccos{\frac{\bm{n}_{\rm s}\!\cdot\!\bm{x}_{\rm s}}{r_{\rm s}}}\!\Big)\!-\!\frac{4m^2}{b^2r_{\rm s}}\!\Big(\!\frac{\bm{n}_{\rm s}\!\cdot\!\bm{x}_{\rm o}}{r_{\rm o}}\!-\!\frac{\bm{n}_{\rm s}\!\cdot\!\bm{x}_{\rm s}}{r_{\rm s}}\!\Big) \\
&&\hskip-.25cm+\frac{15m^2}{4b^2}\!\Big(\frac{\bm{n}_{\rm s}\!\cdot\!\bm{x}_{\rm o}}{r^2_{\rm o}}\!-\!\frac{\bm{n}_{\rm s}\!\cdot\!\bm{x}_{\rm s}}{r_{\rm s}^2}\!\Big)\!-\!\frac{m^2}{2}\!\Big(\frac{\bm{n}_{\rm s}\!\cdot\!\bm{x}_{\rm o}}{r^4_{\rm o}}\!+\!\frac{\bm{n}_{\rm s}\!\cdot\!\bm{x}_{\rm s}}{r_{\rm s}^4}\!\Big)\!+\!\frac{2(\bm{n}_{\rm s}\!\times\!\bm{b})\!\cdot\!\bm{J}}{b^4}\!\bigg[\!\frac{\bm{n}_{\rm s}\!\cdot\!\bm{x}_{\rm o}}{r_{\rm o}}\!-\!\frac{\bm{n}_{\rm s}\!\cdot\!\bm{x}_{\rm s}}{r_{\rm s}} \\
&&\hskip-.25cm+\frac{b^2(\bm{n}_{\rm s}\!\cdot\!\bm{x}_{\rm o})}{r^3_{\rm o}}\!\bigg]\!\bigg\}\bm{b}\!\cdot\!\bm{v}_{\rm o} \!+\!\bigg[\!\frac{2\bm{n}_{\rm s}\!\cdot\!\bm{J}}{r_{\rm s}^3} \!+\! \frac{2\bm{b}\!\cdot\!\bm{J}}{b^2}\!\bigg(\!\frac{\bm{n}_{\rm s}\!\cdot\!\bm{x}_{\rm o}}{b^2 r_{\rm o}} \!-\!\frac{\bm{n}_{\rm s}\!\cdot\!\bm{x}_{\rm s}}{b^2r_{\rm s}}\!-\!\frac{\bm{n}_{\rm s}\!\cdot\!\bm{x}_{\rm s}}{r_{\rm s}^3}\!\bigg)\!\bigg]\!(\bm{n}_{\rm s}\!\times\!\bm{b})\!\cdot\!\bm{v}_{\rm o}~,\\
 B &=& 1\!-\!\bigg[1\!+\!\frac{2m}{r_{\rm s}}\!+\!\frac{2m^2}{r_{\rm s}^2}\!-\!\frac{m^2b^2}{2r_{\rm s}^4}\!+\!\frac{2(\!\bm{n}_{\rm s}\!\times\!\bm{b}\!)\!\cdot\!\bm{J}}{r_{\rm s}^3}\bigg]\!\bm{n}_{\rm s}\!\cdot\!\bm{v}_{\rm s}\!-\!\bigg[\!\frac{m^2}{r_{\rm s}^3}\!-\!\frac{2(\bm{n}_{\rm s}\!\times\!\bm{b})\!\cdot\!\bm{J}}{b^2 r^2_{\rm s}}\!\bigg]\!\frac{\bm{n}_{\rm s}\!\cdot\!\bm{x}_{\rm s}}{r_{\rm s}}\bm{b}\!\cdot\!\bm{v}_{\rm s} \\
&&\hskip-.25cm+\bigg[\!\frac{2\bm{n}_{\rm s}\!\cdot\!\bm{J}}{r_{\rm s}^3} \!-\! \frac{2\bm{b}\!\cdot\!\bm{J}}{b^2}\!\frac{\bm{n}_{\rm s}\!\cdot\!\bm{x}_{\rm s}}{r_{\rm s}^3}\!\bigg](\bm{n}_{\rm s}\!\times\!\bm{b})\!\cdot\!\bm{v}_{\rm s}~.\quad
\end{eqnarray*}
Substituting Eqs. \eqref{eq:coordinate_clock_ratio} and \eqref{eq:derivative-time} into Eq. \eqref{eq:FrequecyShift}, we can obtain, to the 2PM order,
{\small\begin{eqnarray}
\frac{\nu_{\rm o}}{\nu_{\rm s}} &=&\frac{\gamma_{\rm o}}{\gamma_{\rm s}}\frac{1\!-\!\bm{n}_{\rm s}\!\cdot\!\bm{v}_{\rm o}}{1\!-\!\bm{n}_{\rm s}\!\cdot\!\bm{v}_{\rm s}}\! \Bigg\{\!1\!+\!\bigg(\!\frac{1\!+\!\bm{n}_{\rm s}\!\cdot\!\bm{v}_{\rm s}}{1\!-\!\bm{n}_{\rm s}\!\cdot\!\bm{v}_{\rm s}} \!-\! 2\gamma_{\rm s}^2 \bigg)\!\frac{m}{r_{\rm s}}\!-\!\bigg(\!\frac{1\!+\!\bm{n}_{\rm s}\!\cdot\!\bm{v}_{\rm o}}{1\!-\!\bm{n}_{\rm s}\!\cdot\!\bm{v}_{\rm o}} \!-\! 2\gamma_{\rm o}^2 \bigg)\!\frac{m}{r_{\rm o}}\!+\!\frac{2\bm{b}\!\cdot\!\bm{v}_{\rm o}}{1\!-\!\bm{n}_{\rm s}\!\cdot\!\bm{v}_{\rm o}}\frac{m}{b^2}\!\bigg(\!\frac{\bm{n}_{\rm s}\!\cdot\bm{x}_{\rm o}}{r_{\rm o}}\!-\!\frac{\bm{n}_{\rm s}\!\cdot\bm{x}_{\rm s}}{r_{\rm s}}\!\bigg) \nn\\
&&\hskip-.35cm+\!\bigg[\!1\!-\!\frac{13\gamma_{\rm o}^2}{2}\!+\!6\gamma_{\rm o}^4\!+\!\frac{\gamma_{\rm o}^2(\bm{x}_{\rm o}\!\cdot\bm{v}_{\rm o})^2}{2r_{\rm o}^2}\!-\!\bigg(\!2\!+\!4\gamma_{\rm o}^2\!-\!\frac{b^2}{2r_{\rm o}^2}\!\bigg)\!\frac{\bm{n}_{\rm s}\!\cdot\!\bm{v}_{\rm o}}{1\!-\!\bm{n}_{\rm s}\!\cdot\!\bm{v}_{\rm o}}\bigg]\!\frac{m^2}{r_{\rm o}^2}\!+\!\bigg[\frac{5\gamma_{\rm s}^2}{2}\!-\!2\gamma_{\rm s}^4\!-\!\frac{\gamma_{\rm s}^2(\bm{x}_{\rm s}\!\cdot\!\bm{v}_{\rm s})^2}{2r_{\rm s}^2}\!-\!\frac{2\bm{n}_{\rm s}\!\cdot\!\bm{v}_{\rm o}}{1\!-\!\bm{n}_{\rm s}\!\cdot\!\bm{v}_{\rm o}} \nn\\
&&\hskip-.35cm+\frac{\bm{n}_{\rm s}\!\cdot\!\bm{v}_{\rm s}}{1\!-\!\bm{n}_{\rm s}\!\cdot\!\bm{v}_{\rm s}}\!\bigg(\frac{4}{1\!-\!\bm{n}_{\rm s}\!\cdot\!\bm{v}_{\rm s}}\!-\!4\gamma_{\rm s}^2\!-\!\frac{b^2}{2r_{\rm s}^2}\bigg)\!+\!\frac{\bm{b}\!\cdot\!\bm{v}_{\rm s}}{1\!-\!\bm{n}_{\rm s}\!\cdot\!\bm{v}_{\rm s}}\frac{\bm{n}_{\rm s}\!\cdot\!\bm{x}_{\rm s}}{r_{\rm s}^2}\bigg]\!\frac{m^2}{r_{\rm s}^2}\!+\!\bigg[\!(1\!-\!2\gamma_{\rm s}^2)\!\bigg(\!2\gamma_{\rm o}^2\!-\!\frac{1\!+\!\bm{n}_{\rm s}\!\cdot\!\bm{v}_{\rm o}}{1\!-\!\bm{n}_{\rm s}\!\cdot\!\bm{v}_{\rm o}}\bigg)\!-\!\bigg(\!2\!-\!4\gamma_{\rm o}^2 \nn\\
&&\hskip-.35cm+\frac{4\bm{n}_{\rm s}\!\cdot\!\bm{v}_{\rm o}}{1\!-\!\bm{n}_{\rm s}\!\cdot\!\bm{v}_{\rm o}}\bigg)\frac{\bm{n}_{\rm s}\!\cdot\!\bm{v}_{\rm s}}{1\!-\!\bm{n}_{\rm s}\!\cdot\!\bm{v}_{\rm s}}\bigg]\!\frac{m^2}{r_{\rm o}r_{\rm s}}\!+\!\frac{4\bm{n}_{\rm s}\!\cdot\!\bm{v}_{\rm o}}{1\!-\!\bm{n}_{\rm s}\!\cdot\!\bm{v}_{\rm o}}\!\frac{m^2}{b^2}\!\bigg(\!1\!-\!\frac{\bm{n}_{\rm s}\!\cdot\!\bm{x}_{\rm s}}{r_{\rm s}}\frac{\bm{n}_{\rm s}\!\cdot\!\bm{x}_{\rm o}}{r_{\rm o}}\!\bigg)\!+\!\frac{2\gamma_{\rm o}^2(\bm{x}_{\rm o}\!\times\!\bm{J})\!\cdot\!\bm{v}_{\rm o}}{r_{\rm o}^3}\!-\!\frac{2\gamma_{\rm s}^2(\bm{x}_{\rm s}\!\times\!\bm{J})\!\cdot\!\bm{v}_{\rm s}}{r_{\rm s}^3} \nn\\
&&\hskip-.35cm-\frac{2\bm{n}_{\rm s}\!\cdot\!\bm{v}_{\rm o}}{1\!-\!\bm{n}_{\rm s}\!\cdot\!\bm{v}_{\rm o}}\frac{(\bm{n}_{\rm s}\!\times\!\bm{b})\!\cdot\!\bm{J}}{r_{\rm o}^3}\!+\!\bigg(\!1\!-\!\frac{\bm{b}\!\cdot\!\bm{v}_{\rm s}}{\bm{n}_{\rm s}\!\cdot\!\bm{v}_{\rm s}}\frac{\bm{n}_{\rm s}\!\cdot\!\bm{x}_{\rm s}}{b^2}\!\bigg)\frac{2\bm{n}_{\rm s}\!\cdot\!\bm{v}_{\rm s}}{1\!-\!\bm{n}_{\rm s}\!\cdot\!\bm{v}_{\rm s}}\frac{(\bm{n}_{\rm s}\!\times\!\bm{b})\!\cdot\!\bm{J}}{r_{\rm s}^3}\!+\!\frac{2(\bm{n}_{\rm s}\!\times\bm{b})\!\cdot\!\bm{v}_{\rm s}}{1\!-\!\bm{n}_{\rm s}\!\cdot\bm{v}_{\rm s}}\!\bigg(\!\frac{\bm{b}\!\cdot\!\bm{J}}{b^2}\frac{\bm{n}_{\rm s}\!\cdot\bm{x}_{\rm s}}{r_{\rm s}^3}\!-\!\frac{\bm{n}_{\rm s}\!\cdot\!\bm{J}}{r_{\rm s}^3}\bigg) \nn\\
&&\hskip-.35cm-\frac{\bm{b}\!\cdot\!\bm{v}_{\rm o}}{1\!-\!\bm{n}_{\rm s}\!\cdot\!\bm{v}_{\rm o}}\Bigg[\frac{15m^2}{4b^3}\Big(\!\!\arccos{\frac{\bm{n}_{\rm s}\!\cdot\!\bm{x}_{\rm o}}{r_{\rm o}}}\!-\!\arccos{\frac{\bm{n}_{\rm s}\!\cdot\!\bm{x}_{\rm s}}{r_{\rm s}}}\!\Big)\!-\!\frac{15m^2}{4b^2}\!\Big(\frac{\bm{n}_{\rm s}\!\cdot\!\bm{x}_{\rm o}}{r^2_{\rm o}}\!-\!\frac{\bm{n}_{\rm s}\!\cdot\!\bm{x}_{\rm s}}{r_{\rm s}^2}\Big)\!+\!\frac{m^2}{2}\Big(\frac{\bm{n}_{\rm s}\!\cdot\!\bm{x}_{\rm o}}{r^4_{\rm o}}\!+\!\frac{\bm{n}_{\rm s}\!\cdot\!\bm{x}_{\rm s}}{r_{\rm s}^4}\Big) \nn\\
&&\hskip-.35cm+\frac{2m}{b^2}\!\Big[\!(1\!-\!2\gamma_{\rm o}^2)\frac{m}{r_{\rm o}}\!+\!\Big(\!\frac{1\!-\!3\bm{n}_{\rm s}\!\cdot\!\bm{v}_{\rm s}}{1\!-\!\bm{n}_{\rm s}\!\cdot\!\bm{v}_{\rm s}}\!+\!2\gamma_{\rm s}^2\!\Big)\!\frac{m}{r_{\rm s}}\Big]\!\Big(\!\frac{\bm{n}_{\rm s}\!\cdot\!\bm{x}_{\rm o}}{r_{\rm o}}\!-\!\frac{\bm{n}_{\rm s}\!\cdot\!\bm{x}_{\rm s}}{r_{\rm s}}\!\Big)\!-\!\frac{2(\bm{n}_{\rm s}\!\times\!\bm{b})\!\cdot\!\bm{J}}{b^4}\bigg(\!\frac{r_{\rm o}^2\!+\!b^2}{r_{\rm o}^2}\frac{\bm{n}_{\rm s}\!\cdot\!\bm{x}_{\rm o}}{r_{\rm o}}\!-\!\frac{\bm{n}_{\rm s}\!\cdot\!\bm{x}_{\rm s}}{r_{\rm s}}\bigg)\!\Bigg] \nn\\
&&\hskip-.35cm-\frac{2\bm{b}\!\cdot\!\bm{J}}{b^2}\frac{(\bm{n}_{\rm s}\!\times\!\bm{b})\!\cdot\!\bm{v}_{\rm o}}{1\!-\!\bm{n}_{\rm s}\!\cdot\!\bm{v}_{\rm o}}\bigg[\frac{\bm{n}_{\rm s}\!\cdot\!\bm{x}_{\rm s}}{r_{\rm s}^3}\!-\!\frac{1}{b^2}\Big(\!\frac{\bm{n}_{\rm s}\!\cdot\!\bm{x}_{\rm o}}{r_{\rm o}}\!-\!\frac{\bm{n}_{\rm s}\!\cdot\!\bm{x}_{\rm s}}{r_{\rm s}}\!\Big)\!\bigg]\!+\!\frac{2\bm{n}_{\rm s}\!\cdot\!\bm{J}}{r_{\rm s}^3}\frac{(\bm{n}_{\rm s}\!\times\!\bm{b})\!\cdot\!\bm{v}_{\rm o}}{1\!-\!\bm{n}_{\rm s}\!\cdot\!\bm{v}_{\rm o}}\!\Bigg\}~.\label{eq:FrequecyShift-2PM}
\end{eqnarray}}
\noindent It can be seen that this equation reduces to that in special relativity when the gravitational source is absent.

\section{frequency shift in terms of energy}\label{sec:frequencyEnergy}
\begin{figure}[b]
  \centering
  \includegraphics[width=17cm]{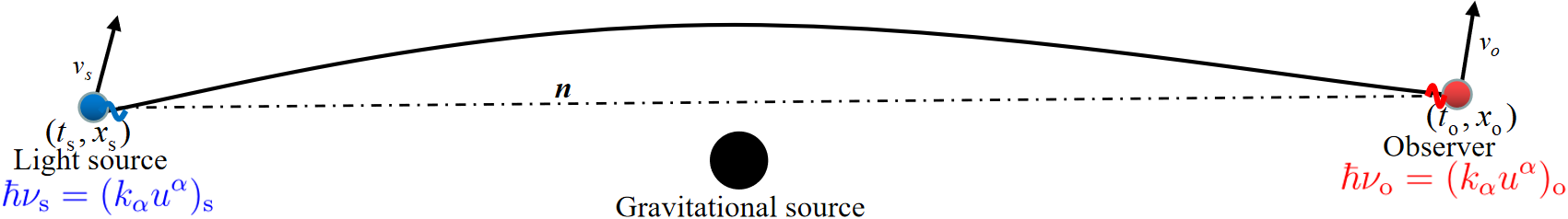}\\
  \caption{Spacetime sketch of the configuration. The gravitational source is stationary, the light source and the observer move with 4-velocity $u^\alpha_{\sss\rm s}=u^0_{\sss\rm s}(1, \bm{v}_{\sss\rm s})$, $u^\alpha_{\sss\rm o}=u^0_{\sss\rm o}(1, \bm{v}_{\sss\rm o})$, respectively. The energy of photon emitting is $\hbar \nu_{\sss\rm s} = (k_\alpha u^\alpha)_{\sss\rm s}$, and the receiving photon's is $\hbar \nu_{\sss\rm o} = (k_\alpha u^\alpha)_{\sss\rm o}$.}\label{fig2}
\end{figure}
There is another method using the definition of the frequency shift in terms of energy\cite{Synge1971}, when one compares the energy of the photon at the points of emission and reception of light, to calculate the frequency shift. The frequency shift in terms of energy is given by
\begin{equation}\label{eq:FrequecyShiftenergy}
  \frac{\nu_{\rm o}}{\nu_{\rm s}}-1=\frac{(k_{{\rm}\alpha} u_{\rm}^\alpha)_{\rm o}}{(k_{\alpha} u^\alpha)_{\rm s}}-1~,
\end{equation}
where $u_{\rm s}^\alpha$, $u_{\rm o}^\alpha$ and $k_{\rm s}^\alpha$, $k_{\rm o}^\alpha$ are the 4-velocities of the emission and reception of light, the wave vectors of photon at the points of emission and reception respectively (see FIG.\ref{fig2}). The wave vector satisfies the propagation equations
\begin{equation}\label{eq:0wavevector}
  k_\mu k^\mu =0~,
\end{equation}
\begin{equation}\label{eq:iwavevector}
  \frac{dk_\mu}{d\lambda}=\frac{1}{2}g_{\alpha\beta,\mu}k^\alpha k^\beta~,
\end{equation}
where $\lambda$ is an affine parameter along the photon trajectory. One can recast $i$-component of Eq.\eqref{eq:iwavevector} as follows
\begin{equation}\label{eq:iwavevector1}
  \frac{dk_i}{dt}=\frac{1}{2} g_{00,i}k^0+g_{0j,i}k^j+\frac{1}{2k^0}g_{ab,i}k^a k^b~.
\end{equation}

Considering that the metric is independent of time $t$ and the velocity of photon $v^i=k^i/k^0$, we easily to obtain $k_0=-\omega$ with $\omega$ being the angular frequency of photon, and expand $k^i$ into 2PM accuracy as
\begin{equation}\label{eq:k^i}
  k^i = \omega(n^i + k_{\sss\rm 1PM}^i + k_{\sss\rm 2PM}^i)~,
\end{equation}
where $k_{\sss\rm 1PM}^i$ and $k_{\sss\rm 2PM}^i$ are the 1PM and 2PM corrections about its flat spacetime value, respectively.
It should be pointed out that we utilize the relationship $k_i=g_{i\rho}k^\rho$ instead of the expansion $k_i = \omega(l_i + k_{i\sss\rm 1PM} + k_{i\sss\rm 2PM})$ with $|l_i|=1$ to solve Eqs. \eqref{eq:0wavevector} and \eqref{eq:iwavevector}. In this way, $k_i$ can be valuated by $\bm{n}_{\rm s}$-, $\bm{b}$- and $(\bm{n}_{\rm s}\times\bm{b})$-components. Therefore, using Kerr metric components in Eq. \eqref{eq:metric-2PM}, we obtain
\begin{equation}\label{eq:k_i1}
  k_i=\omega\bigg[n^i_{\rm s}+\frac{2m}{r}n^i_{\rm s} + k_{\sss\rm 1PM}^i + k_{\sss\rm 2PM}^i + \frac{2m}{r}k_{\sss\rm 1PM}^i +\frac{m^2}{r^2}n^i_{\rm s} + \frac{m^2(\bm{n}_{\rm s}\cdot\bm{x})}{r^4}x^i+\frac{2(\bm{x}\times\bm{J})^i}{r^3}\bigg]~,
\end{equation}
\begin{equation}\label{eq:k^0}
  k^0 = \omega\bigg[1+\frac{2m}{r}+\frac{2m^2}{r^2}+\frac{2(\bm{n}_{\rm s}\times\bm{b})\cdot\bm{J}}{r^3}\bigg]~.
\end{equation}

Substituting Eqs. \eqref{eq:metric-2PM}, \eqref{eq:trajectory-2PM-general}, \eqref{eq:k^i}-\eqref{eq:k^0} into Eqs. \eqref{eq:0wavevector} and \eqref{eq:iwavevector1} and only keeping the 1PM terms, we can obtain
\begin{equation}
  n^i_{\rm s} k_{\sss\rm 1PM}^i = 0, \quad \frac{dk_{\sss\rm 1PM}^i}{dt} = -\frac{2m}{r^3}b^i~.
\end{equation}
Integrating the above differential equation along the light ray trajectory, one can obtain
\begin{equation}\label{eq:k_1PM^i}
  k_{\sss\rm 1PM}^i=-\frac{2m}{b^2}\bigg(\frac{\bm{n}_{\rm s}\cdot\bm{x}}{r}-\frac{\bm{n}_{\rm s}\cdot\bm{x}_{\rm s}}{r_{\rm s}}\bigg)b^i~.
\end{equation}
Similarly, substituting Eqs. \eqref{eq:metric-2PM}, \eqref{eq:trajectory-2PM-general}, \eqref{eq:k^i}-\eqref{eq:k^0} and \eqref{eq:k_1PM^i} into Eqs. \eqref{eq:0wavevector} and \eqref{eq:iwavevector1} and keeping the required order, we obtain
\begin{equation}
  n^i_{\rm s} k_{\sss\rm 2PM}^i = \frac{m^2b^2}{2r^4} - \frac{2m^2}{b^2}\bigg(\frac{\bm{n}_{\rm s}\cdot\bm{x}}{r}-\frac{\bm{n}_{\rm s}\cdot\bm{x}_{\rm s}}{r_{\rm s}}\bigg)^2~,
\end{equation}
\begin{eqnarray}
  \frac{dk_{\sss\rm 2PM}^i}{dt} &=& \!-\!\bigg[\frac{4m^2\!}{r^3}\!\bigg(\!\frac{\bm{n}_{\rm s}\!\cdot\!\bm{x}}{r}\!-\!\frac{\bm{n}_{\rm s}\!\cdot\!\bm{x}_{\rm s}}{r_{\rm s}}\!\bigg)\!+\!\frac{2m^2b^2}{r^5}\frac{\bm{n}_{\rm s}\!\cdot\!\bm{x}}{r}\bigg]\!n^i \!+\! \bigg(\!\frac{4m^2}{r_{\rm s}r^3}\!-\!\frac{4m^2}{b^2 r^2}\!+\!\frac{2m^2}{r^4}\!-\!\frac{2m^2b^2}{r^6}  \nn\\
  &&+ \frac{4m^2}{b^2 r^2}\frac{\bm{n}_{\rm s}\!\cdot\!\bm{x}_{\rm s}}{r_{\rm s}}\frac{\bm{n}_{\rm s}\!\cdot\!\bm{x}}{r}\!\bigg)b^i\!+\!\frac{6(\bm{n}_{\rm s}\!\cdot\!\bm{x})(\bm{x}\!\times\!\bm{J})^i}{r^5}\!-\!\frac{4(\bm{n}_{\rm s}\!\times\!\bm{J})^i}{r^3} \!-\! \frac{6(\bm{n}_{\rm s}\!\times\!\bm{b})\!\cdot\!\bm{J}}{r^5}x^i~.
\end{eqnarray}
After evaluating several integrals and collecting terms, one can obtain the results, to 2PM accuracy,
\begin{eqnarray}
k_{\sss\rm 2PM}^i &=& \bigg\{\!\frac{m^2b^2}{2r^4} \!-\!\frac{2m^2}{b^2}\!\Big(\!\frac{\bm{n}_{\rm s}\cdot\bm{x}}{r} \!-\! \frac{\bm{n}_{\rm s}\cdot\bm{x}_{\rm s}}{r_{\rm s}}\!\Big)^2\!\bigg\}n^i \nn\\
&& + \bigg\{\!\frac{15m^2}{4b^3}\!\bigg(\!\!\arccos\!{\frac{\!\bm{n}_{\rm s}\!\cdot\!\bm{x}\!}{r}} \!-\! \arccos\!{\frac{\!\bm{n}_{\rm s}\!\cdot\!\bm{x}_{\rm s}\!}{r_{\rm s}}}\!\bigg) \!+\!\Big[\frac{4m^2}{b^2r}+\frac{4m^2}{b^2r_{\rm s}}\!-\!\frac{2(\bm{n}_{\rm s}\!\times\!\bm{b})\!\cdot\!\bm{J}}{b^4}\Big]\!\bigg(\!\frac{\bm{n}_{\rm s}\!\cdot\!\bm{x}}{r} \!-\!\frac{\bm{n}_{\rm s}\!\cdot\!\bm{x}_{\rm s}}{r_{\rm s}}\!\bigg)\nn\\
&&\hskip.5cm -\frac{15m^2}{4b^2}\!\bigg(\!\frac{\bm{n}_{\rm s}\!\cdot\!\bm{x}}{r^2} \!-\!\frac{\bm{n}_{\rm s}\!\cdot\!\bm{x}_{\rm s}}{r_{\rm s}^2}\!\bigg) \!-\! \frac{m^2}{2}\!\bigg(\!\frac{\bm{n}_{\rm s}\!\cdot\!\bm{x}}{r^4} \!-\! \frac{\bm{n}_{\rm s}\!\cdot\!\bm{x}_{\rm s}}{r_{\rm s}^4}\!\bigg) \!\bigg\}b^i \nn\\
&& +\bigg\{\!2\bm{n}_{\rm s}\!\cdot\!\bm{J}\bigg(\!\frac{1}{r^3} \!-\!\frac{1}{r_{\rm s}^3}\!\bigg) \!-\! \frac{2\bm{b}\!\cdot\!\bm{J}}{b^2}\!\bigg(\!\frac{\bm{n}_{\rm s}\!\cdot\!\bm{x}}{b^2r} \!-\!\frac{\bm{n}_{\rm s}\!\cdot\!\bm{x}_{\rm s}}{b^2r_{\rm s}}\!+\!\frac{\bm{n}_{\rm s}\!\cdot\!\bm{x}}{r^3} \!-\!\frac{\bm{n}_{\rm s}\!\cdot\!\bm{x}_{\rm s}}{r_{\rm s}^3}\!\bigg)\!\bigg\}(\bm{n}_{\rm s}\!\times\!\bm{b})^i~.\label{eq:ki-2PM}
\end{eqnarray}
Therefore, the covariant wave vector $k_i$ can be got from $k_i=g_{i\rho}k^\rho$ as follows
\begin{eqnarray}
k_i&=&\omega\bigg\{\!\bigg[1\!+\!\frac{2m}{r}\!+\!\frac{4m^2}{r^2}\!-\!\frac{4m^2}{b^2}\!+\!\frac{2m^2}{r_{\rm s}^2}\!-\!\frac{m^2b^2}{2r^4}\!+\!\frac{4m^2}{b^2}\frac{\bm{n}_{\rm s}\!\cdot\!\bm{x}_{\rm s}}{r_{\rm s}}\frac{\bm{n}_{\rm s}\!\cdot\!\bm{x}}{r}\!+\!\frac{2(\bm{n}_{\rm s}\!\times\!\bm{b})\cdot\bm{J}}{r^3}\bigg]n^i \nn\\
&&+\bigg[\!\!-\!\frac{2m}{b^2}\!\Big(\!\frac{\bm{n}_{\rm s}\!\cdot\!\bm{x}}{r}\!-\!\frac{\bm{n}_{\rm s}\!\cdot\!\bm{x}_{\rm s}}{r_{\rm s}}\!\Big)\!+\!\frac{15m^2}{4b^3}\!\Big(\!\!\arccos{\frac{\bm{n}_{\rm s}\!\cdot\!\bm{x}}{r}}\!-\!\arccos{\frac{\bm{n}_{\rm s}\!\cdot\!\bm{x}_{\rm s}}{r_{\rm s}}}\!\Big)\!+\!\frac{4m^2}{b^2r_{\rm s}}\!\Big(\!\frac{\bm{n}_{\rm s}\!\cdot\!\bm{x}}{r}\!-\!\frac{\bm{n}_{\rm s}\!\cdot\!\bm{x}_{\rm s}}{r_{\rm s}}\!\Big) \nn\\
&&-\frac{15m^2}{4b^2}\!\Big(\!\frac{\bm{n}_{\rm s}\!\cdot\!\bm{x}}{r^2}\!-\!\frac{\bm{n}_{\rm s}\!\cdot\!\bm{x}_{\rm s}}{r_{\rm s}^2}\!\Big)\!+\!\frac{m^2}{2}\!\Big(\!\frac{\bm{n}_{\rm s}\!\cdot\!\bm{x}}{r^4}\!+\!\frac{\bm{n}_{\rm s}\!\cdot\!\bm{x}_{\rm s}}{r_{\rm s}^4}\!\Big)\!-\!\frac{2(\bm{n}_{\rm s}\!\times\!\bm{b})\!\cdot\!\bm{J}}{b^4}\!\bigg(\!\frac{\bm{n}_{\rm s}\!\cdot\!\bm{x}}{r}\!-\!\frac{\bm{n}_{\rm s}\!\cdot\!\bm{x}_{\rm s}}{r_{\rm s}}\!+\!\frac{b^2(\bm{n}_{\rm s}\!\cdot\!\bm{x})}{r^3}\!\bigg)\!\bigg]b^i \nn\\
&& -\bigg[\!\frac{2\bm{n}_{\rm s}\!\cdot\!\bm{J}}{r_{\rm s}^3} \!+\! \frac{2\bm{b}\!\cdot\!\bm{J}}{b^2}\!\bigg(\!\frac{\bm{n}_{\rm s}\!\cdot\!\bm{x}}{b^2r} \!-\!\frac{\bm{n}_{\rm s}\!\cdot\!\bm{x}_{\rm s}}{b^2r_{\rm s}}\!-\!\frac{\bm{n}_{\rm s}\!\cdot\!\bm{x}_{\rm s}}{r_{\rm s}^3}\!\bigg)\!\bigg](\bm{n}_{\rm s}\!\times\!\bm{b})^i\bigg\}~.\label{eq:k_i}
\end{eqnarray}
Then the wave vectors $(k_i)_{\rm s}$ of photon at the point of emission is
\begin{eqnarray}
(k_i)_{\rm s}&=&\omega\bigg\{\!\bigg[1\!+\!\frac{2m}{r_{\rm s}}\!+\!\frac{2m^2}{r_{\rm s}^2}\!-\!\frac{m^2b^2}{2r^4_{\rm s}}\!+\!\frac{2(\bm{n}_{\rm s}\!\times\!\bm{b})\cdot\bm{J}}{r^3_{\rm s}}\bigg]n^i -\frac{2(\bm{n}_{\rm s}\!\times\!\bm{b})\!\cdot\!\bm{J}}{b^2}\frac{(\bm{n}_{\rm s}\!\cdot\!\bm{x}_{\rm s})}{r^3_{\rm s}}b^i \nn\\
&&-\bigg[\!\frac{2\bm{n}_{\rm s}\!\cdot\!\bm{J}}{r_{\rm s}^3} \!-\! \frac{2\bm{b}\!\cdot\!\bm{J}}{b^2}\frac{\bm{n}_{\rm s}\!\cdot\!\bm{x}_{\rm s}}{r_{\rm s}^3}\bigg](\bm{n}_{\rm s}\!\times\!\bm{b})^i\bigg\}~.\label{eq:k_i}
\end{eqnarray}
Inserting Eqs.\eqref{eq:coordinate_clock_ratio} and \eqref{eq:k_i} into Eq.\eqref{eq:FrequecyShiftenergy}, we can obtain the frequency shift to 2PM accuracy,  which is exactly same as given by \eqref{eq:FrequecyShift-2PM}.

\section{Some specific scenarios}\label{sec:somecases}

In this section, we first consider the Schwarzschild spacetime, and then explore two important astronomical scenarios where the frequency shift formulas derived in the previous section can be applied.

Scenario 1: When the gravitational source reduces to the Schwarzschild black hole, neglecting the effects of the gravitational rotation, the ratio of frequencies can be expressed as
{\small\begin{eqnarray}
\frac{\nu_{\rm o}}{\nu_{\rm s}} &=&\frac{\gamma_{\rm o}}{\gamma_{\rm s}}\frac{1\!-\!\bm{n}_{\rm s}\!\cdot\!\bm{v}_{\rm o}}{1\!-\!\bm{n}_{\rm s}\!\cdot\!\bm{v}_{\rm s}}\! \Bigg\{\!1\!+\!\bigg(\!\frac{1\!+\!\bm{n}_{\rm s}\!\cdot\!\bm{v}_{\rm s}}{1\!-\!\bm{n}_{\rm s}\!\cdot\!\bm{v}_{\rm s}}\!-\! 2\gamma_{\rm s}^2 \bigg)\!\frac{m}{r_{\rm s}}\!-\!\bigg(\!\frac{1\!+\!\bm{n}_{\rm s}\!\cdot\!\bm{v}_{\rm o}}{1\!-\!\bm{n}_{\rm s}\!\cdot\!\bm{v}_{\rm o}}\!-\! 2\gamma_{\rm o}^2 \bigg)\!\frac{m}{r_{\rm o}}\!+\!\frac{2\bm{b}\!\cdot\!\bm{v}_{\rm o}}{1\!-\!\bm{n}_{\rm s}\!\cdot\!\bm{v}_{\rm o}}\frac{m}{b^2}\!\bigg(\!\frac{\bm{n}_{\rm s}\!\cdot\bm{x}_{\rm o}}{r_{\rm o}}\!-\!\frac{\bm{n}_{\rm s}\!\cdot\bm{x}_{\rm s}}{r_{\rm s}}\!\bigg) \nn\\
&&\hskip-.35cm+\bigg[1\!-\!\frac{13\gamma_{\rm o}^2}{2}\!+\!6\gamma_{\rm o}^4\!+\!\frac{\gamma_{\rm o}^2(\bm{x}_{\rm o}\!\cdot\bm{v}_{\rm o})^2}{2r_{\rm o}^2}\!-\!\bigg(\!2\!+\!4\gamma_{\rm o}^2\!-\!\frac{b^2}{2r_{\rm o}^2}\bigg)\!\frac{\bm{n}_{\rm s}\!\cdot\!\bm{v}_{\rm o}}{1\!-\!\bm{n}_{\rm s}\!\cdot\!\bm{v}_{\rm o}}\bigg]\!\frac{m^2}{r_{\rm o}^2}\!+\!\bigg[\frac{5\gamma_{\rm s}^2}{2}\!-\!2\gamma_{\rm s}^4\!-\!\frac{\gamma_{\rm s}^2(\bm{x}_{\rm s}\!\cdot\!\bm{v}_{\rm s})^2}{2r_{\rm s}^2} \nn\\
&&\hskip-.35cm-\frac{2\bm{n}_{\rm s}\!\cdot\!\bm{v}_{\rm o}}{1\!-\!\bm{n}_{\rm s}\!\cdot\!\bm{v}_{\rm o}}\!+\!\frac{\bm{n}_{\rm s}\!\cdot\!\bm{v}_{\rm s}}{1\!-\!\bm{n}_{\rm s}\!\cdot\!\bm{v}_{\rm s}}\!\bigg(\frac{4}{1\!-\!\bm{n}_{\rm s}\!\cdot\!\bm{v}_{\rm s}}\!-\!4\gamma_{\rm s}^2\!-\!\frac{b^2}{2r_{\rm s}^2}\bigg)\!+\!\frac{\bm{b}\!\cdot\!\bm{v}_{\rm s}}{1\!-\!\bm{n}_{\rm s}\!\cdot\!\bm{v}_{\rm s}}\frac{\bm{n}_{\rm s}\!\cdot\!\bm{x}_{\rm s}}{r_{\rm s}^2}\bigg]\!\frac{m^2}{r_{\rm s}^2}\!+\!\bigg[\!(1\!-\!2\gamma_{\rm s}^2)\!\bigg(\!2\gamma_{\rm o}^2\!-\!\frac{1\!+\!\bm{n}_{\rm s}\!\cdot\!\bm{v}_{\rm o}}{1\!-\!\bm{n}_{\rm s}\!\cdot\!\bm{v}_{\rm o}}\!\bigg) \nn\\
&&\hskip-.35cm-\bigg(\!\frac{2\!+\!2\bm{n}_{\rm s}\!\cdot\!\bm{v}_{\rm o}}{1\!-\!\bm{n}_{\rm s}\!\cdot\!\bm{v}_{\rm o}}\!-\!4\gamma_{\rm o}^2\bigg)\frac{\bm{n}_{\rm s}\!\cdot\!\bm{v}_{\rm s}}{1\!-\!\bm{n}_{\rm s}\!\cdot\!\bm{v}_{\rm s}}\bigg]\!\frac{m^2}{r_{\rm o}r_{\rm s}}\!-\!\frac{\bm{b}\!\cdot\!\bm{v}_{\rm o}}{1\!-\!\bm{n}_{\rm s}\!\cdot\!\bm{v}_{\rm o}}\!\Bigg[\frac{m^2}{2}\!\Big(\frac{\bm{n}_{\rm s}\!\cdot\!\bm{x}_{\rm o}}{r^4_{\rm o}}\!+\!\frac{\bm{n}_{\rm s}\!\cdot\!\bm{x}_{\rm s}}{r_{\rm s}^4}\Big)\!-\!\frac{15m^2}{4b^2}\Big(\!\frac{\bm{n}_{\rm s}\!\cdot\!\bm{x}_{\rm o}}{r^2_{\rm o}}\!-\!\frac{\bm{n}_{\rm s}\!\cdot\!\bm{x}_{\rm s}}{r_{\rm s}^2}\!\Big) \nn\\
&&\hskip-.35cm+\frac{15m^2}{4b^3}\Big(\!\!\arccos\!{\frac{\bm{n}_{\rm s}\!\cdot\!\bm{x}_{\rm o}}{r_{\rm o}}}\!-\!\arccos\!{\frac{\bm{n}_{\rm s}\!\cdot\!\bm{x}_{\rm s}}{r_{\rm s}}}\!\Big)\!+\!\frac{2m}{b^2}\!\Big[\!(1\!-\!2\gamma_{\rm o}^2)\frac{m}{r_{\rm o}}\!+\!\Big(\!\frac{1\!-\!3\bm{n}_{\rm s}\!\cdot\!\bm{v}_{\rm s}}{1\!-\!\bm{n}_{\rm s}\!\cdot\!\bm{v}_{\rm s}}\!+\!2\gamma_{\rm s}^2\!\Big)\!\frac{m}{r_{\rm s}}\Big]\!\Big(\!\frac{\bm{n}_{\rm s}\!\cdot\!\bm{x}_{\rm o}}{r_{\rm o}}\!-\!\frac{\bm{n}_{\rm s}\!\cdot\!\bm{x}_{\rm s}}{r_{\rm s}}\!\Big)\!\Bigg]\nn\\
&&\hskip-.35cm+\frac{4\bm{n}_{\rm s}\!\cdot\!\bm{v}_{\rm o}}{1\!-\!\bm{n}_{\rm s}\!\cdot\!\bm{v}_{\rm o}}\frac{m^2}{b^2}\!\bigg(\!1\!-\!\frac{\bm{n}_{\rm s}\!\cdot\!\bm{x}_{\rm s}}{r_{\rm s}}\frac{\bm{n}_{\rm s}\!\cdot\!\bm{x}_{\rm o}}{r_{\rm o}}\!\bigg)\!\Bigg\}~.\label{eq:FrequecyShift-2PM-Sch}
\end{eqnarray}}
If the velocities of the light source and the observer are non-relativistic, and the ratios of them to the vacuum light speed are small, we can perform an expand the above expression with these ratios, and arrive at the 1PM formulation for the Schwarzschild spacetime given in Ref. \cite{qin2017}.

Scenario 2: We consider the configuration in which the light source is in the vicinity of the gravitational source, and the observer lies in the asymptotically flat region, i.e., $\bm{x}_{\rm o}\rightarrow+\infty$. The ratio of frequencies can be represented as
{\small\begin{eqnarray}
\frac{\nu_{\rm o}}{\nu_{\rm s}} &=&\frac{\gamma_{\rm o}}{\gamma_{\rm s}}\frac{1\!-\!\bm{n}_{\rm s}\!\cdot\!\bm{v}_{\rm o}}{1\!-\!\bm{n}_{\rm s}\!\cdot\!\bm{v}_{\rm s}} \Bigg\{\!1\!+\!\bigg(\!1 \!-\! 2\gamma_{\rm s}^2 \!+\!\frac{2\bm{n}_{\rm s}\!\cdot\!\bm{v}_{\rm s}}{1\!-\!\bm{n}_{\rm s}\!\cdot\!\bm{v}_{\rm s}}\bigg)\!\frac{m}{r_{\rm s}}\!+\!\frac{2\bm{b}\!\cdot\!\bm{v}_{\rm o}}{1\!-\!\bm{n}_{\rm s}\!\cdot\!\bm{v}_{\rm o}}\frac{m}{b^2}\!\bigg(\!1\!-\!\frac{\bm{n}_{\rm s}\!\cdot\bm{x}_{\rm s}}{r_{\rm s}}\!\bigg)\!+\!\bigg[\frac{5\gamma_{\rm s}^2}{2}\!-\!2\gamma_{\rm s}^4\!-\!\frac{\gamma_{\rm s}^2(\bm{x}_{\rm s}\!\cdot\!\bm{v}_{\rm s})^2}{2r_{\rm s}^2} \nn\\
&&\hskip-.35cm-\frac{2\bm{n}_{\rm s}\!\cdot\!\bm{v}_{\rm o}}{1\!-\!\bm{n}_{\rm s}\!\cdot\!\bm{v}_{\rm o}} \!+\!\frac{\bm{n}_{\rm s}\!\cdot\!\bm{v}_{\rm s}}{1\!-\!\bm{n}_{\rm s}\!\cdot\!\bm{v}_{\rm s}}\!\bigg(\frac{4}{1\!-\!\bm{n}_{\rm s}\!\cdot\!\bm{v}_{\rm s}}\!-\!4\gamma_{\rm s}^2\!-\!\frac{b^2}{2r_{\rm s}^2}\bigg)\!+\!\frac{\bm{b}\!\cdot\!\bm{v}_{\rm s}}{1\!-\!\bm{n}_{\rm s}\!\cdot\!\bm{v}_{\rm s}}\frac{\bm{n}_{\rm s}\!\cdot\!\bm{x}_{\rm s}}{r_{\rm s}^2}\bigg]\!\frac{m^2}{r_{\rm s}^2}\!+\!\frac{4\bm{n}_{\rm s}\!\cdot\!\bm{v}_{\rm o}}{1\!-\!\bm{n}_{\rm s}\!\cdot\!\bm{v}_{\rm o}}\frac{m^2}{b^2}\!\bigg(\!1\!-\!\frac{\bm{n}_{\rm s}\!\cdot\!\bm{x}_{\rm s}}{r_{\rm s}}\!\bigg) \nn\\
&&\hskip-.35cm-\frac{2\gamma_{\rm s}^2(\bm{x}_{\rm s}\!\times\!\bm{J})\!\cdot\!\bm{v}_{\rm s}}{r_{\rm s}^3}\!+\!\frac{2\bm{n}_{\rm s}\!\cdot\!\bm{v}_{\rm s}}{1\!-\!\bm{n}_{\rm s}\!\cdot\!\bm{v}_{\rm s}}\bigg(\!1\!-\!\frac{\bm{b}\!\cdot\!\bm{v}_{\rm s}}{\bm{n}_{\rm s}\!\cdot\!\bm{v}_{\rm s}}\frac{\bm{n}_{\rm s}\!\cdot\!\bm{x}_{\rm s}}{b^2}\!\bigg)\frac{(\bm{n}_{\rm s}\!\times\!\bm{b})\!\cdot\!\bm{J}}{r_{\rm s}^3}\!+\!\frac{2(\bm{n}_{\rm s}\!\times\bm{b})\!\cdot\!\bm{v}_{\rm s}}{1\!-\!\bm{n}_{\rm s}\!\cdot\bm{v}_{\rm s}}\!\bigg(\!\frac{\bm{b}\!\cdot\!\bm{J}}{b^2}\frac{\bm{n}_{\rm s}\!\cdot\bm{x}_{\rm s}}{r_{\rm s}^3}\!-\!\frac{\bm{n}_{\rm s}\!\cdot\!\bm{J}}{r_{\rm s}^3}\bigg) \nn\\
&&\hskip-.35cm+\frac{\bm{b}\!\cdot\!\bm{v}_{\rm o}}{1\!-\!\bm{n}_{\rm s}\!\cdot\!\bm{v}_{\rm o}}\!\Bigg[\frac{15m^2}{4b^3}\arccos{\frac{\bm{n}_{\rm s}\!\cdot\!\bm{x}_{\rm s}}{r_{\rm s}}}\!-\!\frac{15m^2}{4b^2}\frac{\bm{n}_{\rm s}\!\cdot\!\bm{x}_{\rm s}}{r_{\rm s}^2}\!-\!\frac{m^2}{2}\frac{\bm{n}_{\rm s}\!\cdot\!\bm{x}_{\rm s}}{r_{\rm s}^4}\!-\!\frac{2m^2}{b^2 r_{\rm s}}\!\Big(\!\frac{1\!-\!3\bm{n}_{\rm s}\!\cdot\!\bm{v}_{\rm s}}{1\!-\!\bm{n}_{\rm s}\!\cdot\!\bm{v}_{\rm s}}\!+\!2\gamma_{\rm s}^2\!\Big)\!\Big(\!1\!-\!\frac{\bm{n}_{\rm s}\!\cdot\!\bm{x}_{\rm s}}{r_{\rm s}}\!\Big) \nn\\
&&\hskip-.35cm+\frac{2(\bm{n}_{\rm s}\!\times\!\bm{b})\!\cdot\!\bm{J}}{b^4}\bigg(\!1\!-\!\frac{\bm{n}_{\rm s}\!\cdot\!\bm{x}_{\rm s}}{r_{\rm s}}\bigg)\!\Bigg]\!-\!\frac{2\bm{b}\!\cdot\!\bm{J}}{b^2}\frac{(\bm{n}_{\rm s}\!\times\!\bm{b})\!\cdot\!\bm{v}_{\rm o}}{1\!-\!\bm{n}_{\rm s}\!\cdot\!\bm{v}_{\rm o}}\bigg[\frac{\bm{n}_{\rm s}\!\cdot\!\bm{x}_{\rm s}}{r_{\rm s}^3}\!-\!\frac{1}{b^2}\Big(\!1\!-\!\frac{\bm{n}_{\rm s}\!\cdot\!\bm{x}_{\rm s}}{r_{\rm s}}\!\Big)\!\bigg]\!+\!\frac{2\bm{n}_{\rm s}\!\cdot\!\bm{J}}{r_{\rm s}^3}\frac{(\bm{n}_{\rm s}\!\times\!\bm{b})\!\cdot\!\bm{v}_{\rm o}}{1\!-\!\bm{n}_{\rm s}\!\cdot\!\bm{v}_{\rm o}}\!\Bigg\}\,.\qquad\label{eq:FrequecyShift-2PM-ro-infty}
\end{eqnarray}}

Scenario 3: When both the light source and the observer are located in the asymptotically flat region, i.e., $\bm{x}_{\rm o}\rightarrow+\infty$ and $\bm{x}_{\rm s}\rightarrow-\infty$, the ratio of frequencies reduces to
%\begin{equation}
%  \frac{\nu_{\rm o}}{\nu_{\rm s}} = \frac{\sqrt{1 \!-\! \bm{v}_{\rm s}^2}}{\sqrt{1 \!-\! \bm{v}_{\rm o}^2}}\frac{1\!-\!\Big(\!1\!-\!\frac{8m^2}{b^2}\!\Big)\bm{n}_{\rm s}\!\cdot\!\bm{v}_{\rm o}\!+\!\Big(\!\frac{4m}{b^2}\!+\!\frac{15\pi m^2}{4b^3}\!+\!\frac{4(\!\bm{n}_{\rm s}\!\times\!\bm{b}\!)\cdot\bm{J}}{b^4}\!\Big)\bm{b}\!\cdot\!\bm{v}_{\rm o}\!+\!\frac{4\bm{b}\cdot\bm{J}}{b^4}(\!\bm{n}_{\rm s}\!\times\!\bm{b}\!)\!\cdot\!\bm{v}_{\rm o}}{1\!-\!\bm{n}_{\rm s}\!\cdot\!\bm{v}_{\rm s}}~.
%\end{equation}
%The above equation can be rewritten as
\begin{equation}\label{eq:infiniteFS}
  \frac{\nu_{\rm o}}{\nu_{\rm s}} = \frac{\gamma_{\rm o}}{\gamma_{\rm s}}\frac{1 - \bm{c}_{\rm o}\cdot\bm{v}_{\rm o}}{1 - \bm{c}_{\rm s}\cdot\bm{v}_{\rm s}}~,
\end{equation}
where
\begin{eqnarray*}
&& \bm{c}_{\rm s }=\bm{n}_{\rm s}~,\\
&&  \bm{c}_{\rm o} = \Big(1-\frac{8m^2}{b^2}\Big)\bm{n}_{\rm s}-\Big[\frac{4m}{b^2}+\frac{15\pi m^2}{4b^3}+\frac{4(\bm{n}_{\rm s}\times\bm{b})\cdot\bm{J}}{b^4}\Big]\bm{b}-\frac{4\bm{b}\cdot\bm{J}}{b^4}(\bm{n}_{\rm s}\times\bm{b})~,
\end{eqnarray*}
are the corresponding velocities of the photon at emission and reception points in the asymptotically flat region, respectively. The derivation of $\bm{c}_{\rm o}$ can be found in Refs.\cite{JiangLin2018,YangJiangLin2019}. The first term of Eq. \eqref{eq:infiniteFS} is the effect due to Lorentz factor, and the second term can be regarded as the Doppler effect. We can use this formula to estimate the mass and angular momentum of the rotating gravitational source.

\section{Summary}\label{sec:summary}
In this work, we have investigated the frequency shift of light emitted from a moving source and received by a moving observer in the presence of a gravitational field generated by a rotating body. We employ two approaches to derive the frequency shift formulas up to the 2PM approximation, considering the combined effects of gravitation and kinematics. The first approach involves comparing the time interval between two neighboring crests of the light ray at the emission and reception points, while the second approach calculates the ratio of the photon energy at two different positions. We have demonstrated that these two methods yield the same results up to the 2PM approximation. Furthermore, we explore two interesting scenarios: the first one is that the moving light source is near to the gravitational source and the moving observer is in the asymptotically flat region, and the other is both the light source and the observer are in the asymptotically flat region. For the latter case, a very concise formula is achieved.

\section*{ACKNOWLEDGEMENT}
This work was supported in part by the National Natural Science Foundation of China (Grant Nos. 11973025 and 12247157 and 12205139), Scientific Research Fund of Hunan Provincial Education Department (Grant No. 22B0446), and Natural Science Foundation of Hunan Province(Grant No. 2022JJ40347).


\begin{thebibliography}{25}%
\makeatletter
\providecommand \@ifxundefined [1]{%
 \@ifx{#1\undefined}
}%
\providecommand \@ifnum [1]{%
 \ifnum #1\expandafter \@firstoftwo
 \else \expandafter \@secondoftwo
 \fi
}%
\providecommand \@ifx [1]{%
 \ifx #1\expandafter \@firstoftwo
 \else \expandafter \@secondoftwo
 \fi
}%
\providecommand \natexlab [1]{#1}%
\providecommand \enquote  [1]{``#1''}%
\providecommand \bibnamefont  [1]{#1}%
\providecommand \bibfnamefont [1]{#1}%
\providecommand \citenamefont [1]{#1}%
\providecommand \href@noop [0]{\@secondoftwo}%
\providecommand \href [0]{\begingroup \@sanitize@url \@href}%
\providecommand \@href[1]{\@@startlink{#1}\@@href}%
\providecommand \@@href[1]{\endgroup#1\@@endlink}%
\providecommand \@sanitize@url [0]{\catcode `\\12\catcode `\$12\catcode
  `\&12\catcode `\#12\catcode `\^12\catcode `\_12\catcode `\%12\relax}%
\providecommand \@@startlink[1]{}%
\providecommand \@@endlink[0]{}%
\providecommand \url  [0]{\begingroup\@sanitize@url \@url }%
\providecommand \@url [1]{\endgroup\@href {#1}{\urlprefix }}%
\providecommand \urlprefix  [0]{URL }%
\providecommand \Eprint [0]{\href }%
\providecommand \doibase [0]{http://dx.doi.org/}%
\providecommand \selectlanguage [0]{\@gobble}%
\providecommand \bibinfo  [0]{\@secondoftwo}%
\providecommand \bibfield  [0]{\@secondoftwo}%
\providecommand \translation [1]{[#1]}%
\providecommand \BibitemOpen [0]{}%
\providecommand \bibitemStop [0]{}%
\providecommand \bibitemNoStop [0]{.\EOS\space}%
\providecommand \EOS [0]{\spacefactor3000\relax}%
\providecommand \BibitemShut  [1]{\csname bibitem#1\endcsname}%
\let\auto@bib@innerbib\@empty
%</preamble>
\bibitem [{\citenamefont {Adams}(1925)}]{Adams1925}%
  \BibitemOpen
  \bibfield  {author} {\bibinfo {author} {\bibfnamefont {W.~S.}\ \bibnamefont
  {Adams}},\ }\href@noop {} {\bibfield  {journal} {\bibinfo  {journal} {Proc.
  Nat. Acad. Sci.}\ }\textbf {\bibinfo {volume} {11}},\ \bibinfo {pages} {382}
  (\bibinfo {year} {1925})}\BibitemShut {NoStop}%
\bibitem [{\citenamefont {Pound}\ and\ \citenamefont
  {Rebka}(1959)}]{Pound-Rebka1959}%
  \BibitemOpen
  \bibfield  {author} {\bibinfo {author} {\bibfnamefont {R.~V.}\ \bibnamefont
  {Pound}}\ and\ \bibinfo {author} {\bibfnamefont {G.~A.}\ \bibnamefont
  {Rebka}},\ }\href@noop {} {\bibfield  {journal} {\bibinfo  {journal} {Phys.
  Rev. Lett.}\ }\textbf {\bibinfo {volume} {3}},\ \bibinfo {pages} {439}
  (\bibinfo {year} {1959})}\BibitemShut {NoStop}%
\bibitem [{\citenamefont {Vessot}\ and\ \citenamefont
  {Levine}(1979)}]{Vessot1979}%
  \BibitemOpen
  \bibfield  {author} {\bibinfo {author} {\bibfnamefont {R.~F.~C.}\
  \bibnamefont {Vessot}}\ and\ \bibinfo {author} {\bibfnamefont {M.~W.}\
  \bibnamefont {Levine}},\ }\href@noop {} {\bibfield  {journal} {\bibinfo
  {journal} {Gen. Relativ. Gravit.}\ }\textbf {\bibinfo {volume} {10}},\
  \bibinfo {pages} {181} (\bibinfo {year} {1979})}\BibitemShut {NoStop}%
\bibitem [{\citenamefont {Vessot}\ \emph {et~al.}(1980)\citenamefont {Vessot},
  \citenamefont {Levine}, \citenamefont {Mattison}, \citenamefont {Blomberg},
  \citenamefont {Hoffman}, \citenamefont {Nystrom}, \citenamefont {Farrel},
  \citenamefont {Decher}, \citenamefont {Eby}, \citenamefont {Baugher},
  \citenamefont {Watts}, \citenamefont {Teuber},\ and\ \citenamefont
  {Wills}}]{Vessot1980}%
  \BibitemOpen
  \bibfield  {author} {\bibinfo {author} {\bibfnamefont {R.~F.~C.}\
  \bibnamefont {Vessot}}, \bibinfo {author} {\bibfnamefont {M.~W.}\
  \bibnamefont {Levine}}, \bibinfo {author} {\bibfnamefont {E.~M.}\
  \bibnamefont {Mattison}}, \bibinfo {author} {\bibfnamefont {E.~L.}\
  \bibnamefont {Blomberg}}, \bibinfo {author} {\bibfnamefont {T.~E.}\
  \bibnamefont {Hoffman}}, \bibinfo {author} {\bibfnamefont {G.~U.}\
  \bibnamefont {Nystrom}}, \bibinfo {author} {\bibfnamefont {B.~F.}\
  \bibnamefont {Farrel}}, \bibinfo {author} {\bibfnamefont {R.}~\bibnamefont
  {Decher}}, \bibinfo {author} {\bibfnamefont {P.~B.}\ \bibnamefont {Eby}},
  \bibinfo {author} {\bibfnamefont {C.~R.}\ \bibnamefont {Baugher}}, \bibinfo
  {author} {\bibfnamefont {J.~W.}\ \bibnamefont {Watts}}, \bibinfo {author}
  {\bibfnamefont {D.~L.}\ \bibnamefont {Teuber}}, \ and\ \bibinfo {author}
  {\bibfnamefont {F.~D.}\ \bibnamefont {Wills}},\ }\href {\doibase
  10.1103/PhysRevLett.45.2081} {\bibfield  {journal} {\bibinfo  {journal}
  {Phys. Rev. Lett.}\ }\textbf {\bibinfo {volume} {45}},\ \bibinfo {pages}
  {2081} (\bibinfo {year} {1980})}\BibitemShut {NoStop}%
\bibitem [{\citenamefont {Vessot}(1989)}]{Vessot1989}%
  \BibitemOpen
  \bibfield  {author} {\bibinfo {author} {\bibfnamefont {R.~F.~C.}\
  \bibnamefont {Vessot}},\ }\href@noop {} {\bibfield  {journal} {\bibinfo
  {journal} {Adv. Space Res.}\ }\textbf {\bibinfo {volume} {9}},\ \bibinfo
  {pages} {21} (\bibinfo {year} {1989})}\BibitemShut {NoStop}%
\bibitem [{\citenamefont {M\"{u}ller}\ \emph {et~al.}(2010)\citenamefont
  {M\"{u}ller}, \citenamefont {Peters},\ and\ \citenamefont
  {Chu}}]{Muller2010}%
  \BibitemOpen
  \bibfield  {author} {\bibinfo {author} {\bibfnamefont {H.}~\bibnamefont
  {M\"{u}ller}}, \bibinfo {author} {\bibfnamefont {A.}~\bibnamefont {Peters}},
  \ and\ \bibinfo {author} {\bibfnamefont {S.}~\bibnamefont {Chu}},\
  }\href@noop {} {\bibfield  {journal} {\bibinfo  {journal} {Nature}\ }\textbf
  {\bibinfo {volume} {463}},\ \bibinfo {pages} {926} (\bibinfo {year}
  {2010})}\BibitemShut {NoStop}%
\bibitem [{\citenamefont {Poli}\ \emph {et~al.}(2011)\citenamefont {Poli},
  \citenamefont {Wang}, \citenamefont {Tarallo}, \citenamefont {Alberti},
  \citenamefont {Prevedelli},\ and\ \citenamefont {Tino}}]{Poli2011}%
  \BibitemOpen
  \bibfield  {author} {\bibinfo {author} {\bibfnamefont {N.}~\bibnamefont
  {Poli}}, \bibinfo {author} {\bibfnamefont {F.-Y.}\ \bibnamefont {Wang}},
  \bibinfo {author} {\bibfnamefont {M.~G.}\ \bibnamefont {Tarallo}}, \bibinfo
  {author} {\bibfnamefont {A.}~\bibnamefont {Alberti}}, \bibinfo {author}
  {\bibfnamefont {M.}~\bibnamefont {Prevedelli}}, \ and\ \bibinfo {author}
  {\bibfnamefont {G.~M.}\ \bibnamefont {Tino}},\ }\href {\doibase
  10.1103/PhysRevLett.106.038501} {\bibfield  {journal} {\bibinfo  {journal}
  {Phys. Rev. Lett.}\ }\textbf {\bibinfo {volume} {106}},\ \bibinfo {pages}
  {038501} (\bibinfo {year} {2011})}\BibitemShut {NoStop}%
\bibitem [{\citenamefont {Jenkins}(1969)}]{Jenkins1969}%
  \BibitemOpen
  \bibfield  {author} {\bibinfo {author} {\bibfnamefont {R.~E.}\ \bibnamefont
  {Jenkins}},\ }\href@noop {} {\bibfield  {journal} {\bibinfo  {journal}
  {Astron. J.}\ }\textbf {\bibinfo {volume} {74}},\ \bibinfo {pages} {960}
  (\bibinfo {year} {1969})}\BibitemShut {NoStop}%
\bibitem [{\citenamefont {Ohanian}\ and\ \citenamefont
  {Ruffini}(1994)}]{Ruffini1994}%
  \BibitemOpen
  \bibfield  {author} {\bibinfo {author} {\bibfnamefont {H.}~\bibnamefont
  {Ohanian}}\ and\ \bibinfo {author} {\bibfnamefont {R.}~\bibnamefont
  {Ruffini}},\ }\href@noop {} {\emph {\bibinfo {title} {Gravitation and
  Spacetime}}}\ (\bibinfo  {publisher} {Norton and Company},\ \bibinfo
  {address} {New York},\ \bibinfo {year} {1994})\BibitemShut {NoStop}%
\bibitem [{\citenamefont {Krisher}(1993)}]{Krisher1993}%
  \BibitemOpen
  \bibfield  {author} {\bibinfo {author} {\bibfnamefont {T.~P.}\ \bibnamefont
  {Krisher}},\ }\href@noop {} {\bibfield  {journal} {\bibinfo  {journal} {Phys.
  Rev. D}\ }\textbf {\bibinfo {volume} {48}},\ \bibinfo {pages} {4639}
  (\bibinfo {year} {1993})}\BibitemShut {NoStop}%
\bibitem [{\citenamefont {Kopeikin}\ and\ \citenamefont
  {Sch\"{a}fer}(1999)}]{Kopeikin1999}%
  \BibitemOpen
  \bibfield  {author} {\bibinfo {author} {\bibfnamefont {S.~M.}\ \bibnamefont
  {Kopeikin}}\ and\ \bibinfo {author} {\bibfnamefont {G.}~\bibnamefont
  {Sch\"{a}fer}},\ }\href@noop {} {\bibfield  {journal} {\bibinfo  {journal}
  {Phys. Rev. D}\ }\textbf {\bibinfo {volume} {60}},\ \bibinfo {pages} {124002}
  (\bibinfo {year} {1999})}\BibitemShut {NoStop}%
\bibitem [{\citenamefont {Dubey}\ and\ \citenamefont
  {Sen}(2015{\natexlab{a}})}]{Sen2015a}%
  \BibitemOpen
  \bibfield  {author} {\bibinfo {author} {\bibfnamefont {A.~K.}\ \bibnamefont
  {Dubey}}\ and\ \bibinfo {author} {\bibfnamefont {A.~K.}\ \bibnamefont
  {Sen}},\ }\href@noop {} {\bibfield  {journal} {\bibinfo  {journal} {Int. J.
  Theor. Phys.}\ }\textbf {\bibinfo {volume} {54}},\ \bibinfo {pages} {2398}
  (\bibinfo {year} {2015}{\natexlab{a}})}\BibitemShut {NoStop}%
\bibitem [{\citenamefont {Dubey}\ and\ \citenamefont
  {Sen}(2015{\natexlab{b}})}]{Sen2015b}%
  \BibitemOpen
  \bibfield  {author} {\bibinfo {author} {\bibfnamefont {A.~K.}\ \bibnamefont
  {Dubey}}\ and\ \bibinfo {author} {\bibfnamefont {A.~K.}\ \bibnamefont
  {Sen}},\ }\href@noop {} {\bibfield  {journal} {\bibinfo  {journal}
  {Astrophys. Space Sci.}\ }\textbf {\bibinfo {volume} {360}},\ \bibinfo
  {pages} {29} (\bibinfo {year} {2015}{\natexlab{b}})}\BibitemShut {NoStop}%
\bibitem [{\citenamefont {Deng}(2016)}]{Deng2016}%
  \BibitemOpen
  \bibfield  {author} {\bibinfo {author} {\bibfnamefont {X.-M.}\ \bibnamefont
  {Deng}},\ }\href@noop {} {\bibfield  {journal} {\bibinfo  {journal} {Int. J.
  Mod. Phys. D}\ }\textbf {\bibinfo {volume} {25}},\ \bibinfo {pages} {1650082}
  (\bibinfo {year} {2016})}\BibitemShut {NoStop}%
\bibitem [{\citenamefont {Qin}\ and\ \citenamefont {Shao}(2017)}]{qin2017}%
  \BibitemOpen
  \bibfield  {author} {\bibinfo {author} {\bibfnamefont {C.-G.}\ \bibnamefont
  {Qin}}\ and\ \bibinfo {author} {\bibfnamefont {C.-G.}\ \bibnamefont {Shao}},\
  }\href@noop {} {\bibfield  {journal} {\bibinfo  {journal} {Phys. Rev. D}\
  }\textbf {\bibinfo {volume} {96}},\ \bibinfo {pages} {024003} (\bibinfo
  {year} {2017})}\BibitemShut {NoStop}%
\bibitem [{\citenamefont {Kuntz}\ and\ \citenamefont
  {Leyde}(2023)}]{Kuntz2023}%
  \BibitemOpen
  \bibfield  {author} {\bibinfo {author} {\bibfnamefont {A.}~\bibnamefont
  {Kuntz}}\ and\ \bibinfo {author} {\bibfnamefont {K.}~\bibnamefont {Leyde}},\
  }\href {\doibase 10.1103/PhysRevD.108.024002} {\bibfield  {journal} {\bibinfo
   {journal} {Phys. Rev. D}\ }\textbf {\bibinfo {volume} {108}},\ \bibinfo
  {pages} {024002} (\bibinfo {year} {2023})}\BibitemShut {NoStop}%
\bibitem [{\citenamefont {Jiang}\ and\ \citenamefont
  {Lin}(2014)}]{JiangLin2014}%
  \BibitemOpen
  \bibfield  {author} {\bibinfo {author} {\bibfnamefont {C.}~\bibnamefont
  {Jiang}}\ and\ \bibinfo {author} {\bibfnamefont {W.}~\bibnamefont {Lin}},\
  }\href@noop {} {\bibfield  {journal} {\bibinfo  {journal} {Gen. Relativ.
  Gravit.}\ }\textbf {\bibinfo {volume} {46}},\ \bibinfo {pages} {1671}
  (\bibinfo {year} {2014})}\BibitemShut {NoStop}%
\bibitem [{\citenamefont {Lin}\ and\ \citenamefont
  {Jiang}(2014)}]{LinJiang2014}%
  \BibitemOpen
  \bibfield  {author} {\bibinfo {author} {\bibfnamefont {W.}~\bibnamefont
  {Lin}}\ and\ \bibinfo {author} {\bibfnamefont {C.}~\bibnamefont {Jiang}},\
  }\href@noop {} {\bibfield  {journal} {\bibinfo  {journal} {Phys. Rev. D}\
  }\textbf {\bibinfo {volume} {89}},\ \bibinfo {pages} {087502} (\bibinfo
  {year} {2014})}\BibitemShut {NoStop}%
\bibitem [{\citenamefont {Weinberg}(1972)}]{Weinberg1972}%
  \BibitemOpen
  \bibfield  {author} {\bibinfo {author} {\bibfnamefont {S.}~\bibnamefont
  {Weinberg}},\ }\href@noop {} {\emph {\bibinfo {title} {Gravitation and
  Cosmology: Principles and Applications of the General Theory of
  Relativity}}}\ (\bibinfo  {publisher} {Wiley},\ \bibinfo {address} {New
  York},\ \bibinfo {year} {1972})\BibitemShut {NoStop}%
\bibitem [{\citenamefont {Will}(1981)}]{Will1981}%
  \BibitemOpen
  \bibfield  {author} {\bibinfo {author} {\bibfnamefont {C.~M.}\ \bibnamefont
  {Will}},\ }\href@noop {} {\emph {\bibinfo {title} {Theory and Experiment in
  Gravitational Physics}}}\ (\bibinfo  {publisher} {Cambridge University
  Press},\ \bibinfo {address} {Cambridge},\ \bibinfo {year} {1981})\BibitemShut
  {NoStop}%
\bibitem [{\citenamefont {Kopeikin}\ \emph {et~al.}(2012)\citenamefont
  {Kopeikin}, \citenamefont {Efroimsky},\ and\ \citenamefont
  {Kaplan}}]{KopeikinEfroimskyKaplan2012}%
  \BibitemOpen
  \bibfield  {author} {\bibinfo {author} {\bibfnamefont {S.~M.}\ \bibnamefont
  {Kopeikin}}, \bibinfo {author} {\bibfnamefont {M.}~\bibnamefont {Efroimsky}},
  \ and\ \bibinfo {author} {\bibfnamefont {G.}~\bibnamefont {Kaplan}},\
  }\href@noop {} {\emph {\bibinfo {title} {Relativistic Celestical Mechanics of
  the Solar System}}}\ (\bibinfo  {publisher} {Wiley-VCH},\ \bibinfo {address}
  {New York},\ \bibinfo {year} {2012})\BibitemShut {NoStop}%
\bibitem [{\citenamefont {Jiang}\ and\ \citenamefont
  {Lin}(2018)}]{JiangLin2018}%
  \BibitemOpen
  \bibfield  {author} {\bibinfo {author} {\bibfnamefont {C.}~\bibnamefont
  {Jiang}}\ and\ \bibinfo {author} {\bibfnamefont {W.}~\bibnamefont {Lin}},\
  }\href@noop {} {\bibfield  {journal} {\bibinfo  {journal} {Phys. Rev. D}\
  }\textbf {\bibinfo {volume} {97}},\ \bibinfo {pages} {024045} (\bibinfo
  {year} {2018})}\BibitemShut {NoStop}%
\bibitem [{\citenamefont {Synge}(1971)}]{Synge1971}%
  \BibitemOpen
  \bibfield  {author} {\bibinfo {author} {\bibfnamefont {J.~L.}\ \bibnamefont
  {Synge}},\ }\href@noop {} {\emph {\bibinfo {title} {Relativity: The General
  Theory}}}\ (\bibinfo  {publisher} {North-Holland},\ \bibinfo {address}
  {Amsterdam},\ \bibinfo {year} {1971})\BibitemShut {NoStop}%
\bibitem [{\citenamefont {Brumberg}(1972)}]{Brumberg1972}%
  \BibitemOpen
  \bibfield  {author} {\bibinfo {author} {\bibfnamefont {V.~A.}\ \bibnamefont
  {Brumberg}},\ }\href@noop {} {\emph {\bibinfo {title} {Relativistic Celestial
  Mechanics (in Russian)}}}\ (\bibinfo  {publisher} {Nauka},\ \bibinfo
  {address} {Moskow},\ \bibinfo {year} {1972})\BibitemShut {NoStop}%
\bibitem [{\citenamefont {Yang}\ \emph {et~al.}(2019)\citenamefont {Yang},
  \citenamefont {Jiang},\ and\ \citenamefont {Lin}}]{YangJiangLin2019}%
  \BibitemOpen
  \bibfield  {author} {\bibinfo {author} {\bibfnamefont {B.}~\bibnamefont
  {Yang}}, \bibinfo {author} {\bibfnamefont {C.}~\bibnamefont {Jiang}}, \ and\
  \bibinfo {author} {\bibfnamefont {W.}~\bibnamefont {Lin}},\ }\href {\doibase
  10.1088/1361-6382/ab0ec9} {\bibfield  {journal} {\bibinfo  {journal} {Class.
  Quantum Grav.}\ }\textbf {\bibinfo {volume} {36}},\ \bibinfo {pages} {085010}
  (\bibinfo {year} {2019})}\BibitemShut {NoStop}%
\end{thebibliography}
\end{document}